\documentclass[12pt]{article}
\usepackage{amssymb,amsmath}
\usepackage{graphicx}
\newtheorem{lem}{Lemma}
\newtheorem{prop}{Proposition}
\newtheorem{cor}{Corollary}
\newtheorem{theor}{Theorem}
\newtheorem{rem}{Remark}

\def\curl{\mathop{\rm curl}}

\font\Bbb=msbm10 scaled 1200

\def\dst{\displaystyle}
\def\ola{\overleftarrow}
\def\ora{\overrightarrow}

\def\am{\stackrel {M} {*}}

\def\sf{\mbox{\hspace{1pt}\raisebox{-8pt}{$
    \lefteqn{ {\scriptstyle F} }{\mbox
                {\raisebox{8pt}{$\star$} }
                     }
                                          $}
             \hspace{-6pt}}
       }

\def\sff{\mbox{\hspace{1pt}\raisebox{-8pt}{$
    \lefteqn{ {\scriptstyle F} }{\mbox
                {\raisebox{8pt}{$\ostar$} }
                     }
                                          $}
             \hspace{-6pt}}
       }

\def\amf{\stackrel {\tau}
{\mbox{\hspace{1pt}\raisebox{-8pt}{$
    \lefteqn{ {\scriptstyle F} }{\mbox
                {\raisebox{8pt}{$\star$} }
                     }
                                          $}
             \hspace{-6pt}}  }
       }

\def\ostar{{\odot}}

\def\A{{\cal A}}

\def\H{{\cal H}}
\def\D{{\cal D}}
\def\F{\mathop{\rm Flux}}
\def\R{\mbox{\Bbb R}}

\def\pa{\partial}
\def\p{\partial}
\def\E{{\cal E}}
\def\L{{\cal L}}
\def\U{{\cal U}}

\begin{document}

\title {Symplectic areas,  quantization, and
dynamics in electromagnetic fields}

\author{M. V. Karasev$^{(a,b)}$ and  T. A. Osborn$^{(a)}$
\\
\\
\small
{$^{(a)}$\sl  Department of Physics and Astronomy}\\
\small
{\sl University of Manitoba}\\
\small
{\sl Winnipeg, MB, Canada, R3T 2N2 }\\
\\
\small
{$^{(b)}$\sl Department of Applied Mathematics}\\
\small
{\sl Moscow Institute of Electronics and Mathematics}\\
\small
{\sl Moscow 109028, Russia }
}
\date{}
\maketitle

\begin{abstract}
        A gauge invariant quantization in a closed integral form
is developed over a linear
phase space endowed with an inhomogeneous Faraday electromagnetic
tensor. An analog of the Groenewold product formula
(corresponding to Weyl ordering) is obtained via a membrane
magnetic area, and extended to the product of $N$ symbols. The
problem of ordering in quantization is related to different
configurations of membranes: a choice of configuration
determines a phase factor that fixes the ordering
and controls a symplectic groupoid structure on
the secondary phase space. A gauge invariant solution of the
quantum evolution problem for a charged particle in an
electromagnetic field is represented in
an exact continual form and in the semiclassical
approximation via the area of dynamical membranes.
\end{abstract}

\section{Introduction and Overview}

        The works by Berezin \cite{3}, Berry \cite{5,6}, and Marinov
\cite{32}
have introduced into mathematical physics elegant formulas
representing three primary quantum objects (the Weyl
non-commutative product, semiclassical eigenfunctions,  and the
evolution wave functions) in terms of symplectic area of simple
two-dimensional surfaces (membranes) whose boundary consists of
line segments and pieces of Hamiltonian trajectories
in phase space.  The area of
these membranes is determined with respect to the canonical
2-form
$$
\omega_0 = \frac12 J_{jk} dx^{k}\wedge dx^{j},  \qquad x=(q,p)\in
\R^{2n}, \qquad J=\left[\begin{matrix} 0&I \cr -I&0\end{matrix}\right]\, .
$$

        Attempts to generalize some of these formulas to
phase spaces with a generic symplectic form have
been undertaken.  For symmetric spaces (see \cite{4}) the opportunity to
represent the quantum product via the area of triangle membranes
was mentioned by Berezin, and it was actually proved  in
\cite{52}, that the semiclassical phase of the product-generating
kernel is given by such an area in this case.  Over K\"ahlerian
manifolds formulas for the Wick product and for the solutions of
stationary or evolution problems via the area of some membranes
in the complexification of phase space were
 obtained in \cite{22,23,24}
(see also \cite{25}).
However, formulas which use only the usual symplectic area are
still unavailable for the general case.

        In the present paper we analyze three problems related
to this topic.

        A first and trivial remark, which one can make regarding
generalization of Berezin--Berry--Marinov geometrical picture, is that the
specific matrix $J$ in the definition of
the symplectic form $\omega_0$ can be replaced by an arbitrary
skew-symmetric matrix without any changes in the geometrical
picture. In particular,
one
can take $J$ to be
the matrix $\left[\begin{matrix}  F&I \cr -I&0\end{matrix}\right]$, where the
constant block $F$ represents a homogenous electromagnetic field. But the
next and more interesting generalization to consider is that of
inhomogeneous (not constant) tensors $F$.

        The second natural question is about ordering.  All treatments
of Berezin, Berry and Marinov picture were made for one specific ordering
choice: for the Weyl symmetrization of the noncommutative coordinates
(operators).  What happens with other possible orderings?

        The third question which our paper addresses is the
application of such Weyl and non--Weyl symbolic calculus, in the
presence of
an inhomogeneous field, to solving the Cauchy problem and developing a
semiclassical representation.

We begin with the first question and
consider the linear phase space $\R^{2n}$ with the following
inhomogeneous symplectic form
\begin{equation}
\omega_F = \omega_0 + F, \qquad F = \frac12
F_{jk}(q)\,dq^{k}\wedge dq^{j}.
\end{equation}
In cases $n=3, n=4$ the form (1.1) describes the
structure of the phase space for charged particles in an
electromagnetic field \cite{49,15}; the additional summand $F$ is
the Faraday
2--form multiplied by the charge coupling constant $e/c$.
For simplicity of
notation we include this constant in $F_{jk}$; also
note that the order of indices $j,k$ in (1.1)
is opposite to that used in some textbooks~\cite{48}.

In detail one has the following. In the 3-dimensional case with
$q=(q^1,q^2,q^3)$ then
$$
F_{jk} = \frac{e}{c} \epsilon_{kjl}B^l \qquad (j,k=1,2,3),
$$
$$
F = \frac{e}{c}\Big(B^1(q)\,dq^2\wedge
dq^3 + B^2(q)\,dq^3 \wedge dq^1 + B^3(q)\, dq^1 \wedge dq^2 \Big).
\eqno{(1.2\rm a)}
$$
Here $B$ is the magnetic field, and $dF = 0$ is
equivalent to ${\rm div} B = 0$.  In the 4-dimensional case
with $q=(q^0,q^1,q^2,q^3)$, $q^0\equiv ct$ one has

$$
F_{jk} =
\frac{e}{c} \epsilon_{kjl}B^l,  \qquad (j,k=1,2,3),
\qquad F_{0j} = \frac e c E_j  \qquad (j=1,2,3),
$$
$$
F = \frac{e}{c}\Big(B^1(t,q)\,dq^2\wedge
dq^3 + B^2(t,q)\,dq^3 \wedge dq^1 + B^3(t,q)\, dq^1 \wedge dq^2 \Big)
+
e E_j(t,q)dq^j\wedge dt.
\eqno{(1.2\rm b)}
$$
The electric field is
denoted by $E$,  and $dF=0$ is equivalent to
the pair of Maxwell equations
$c^{-1}{\partial B / {\partial t} } + {\rm{curl}} E =0$,
$\mbox{div}\,B=0$; see \cite{35, 48}.

\setcounter{equation}{2}

The form $\omega_F$ is called the {\em magnetic symplectic form}.
We show how this form generates the Weyl--type associative
product  $\sf$ of functions over the phase space. The $q^j,p_k$
coordinates on
this space correspond to the position of the charged particle
and its gauge invariant kinetic momentum.  The commutation
relations between the corresponding  quantum operators $\hat
q^j, \hat p_k$ are the following
\begin{equation}
[\hat q^j, \hat q^k ] = 0,
\qquad [\hat q^j,\hat p_k] = i\hbar \delta^j_k, \qquad [\hat
p_j, \hat p_k] = i\hbar F_{kj}(\hat q).
\end{equation}

The usual realization of these operators in the Hilbert space $L^2(\R_q^n)$ is
$\hat q = q$, $\hat p= -i\hbar \p/ \p q -(e/c) \Phi(q)$, where
$(e/c) d(\Phi dq)=F$, and $\Phi$ is the gauge potential.
Specifically, for $n=3$:
$$
\Phi=(\A_1,\A_2,\A_3), \qquad {\rm{curl}}\A = B, \qquad
\hat p_j = -i\hbar\frac{\p}{\p q_j} - \frac ec\A_j(q)\quad (j=1,2,3);
\eqno(1.4\rm a)
$$
for $n= 4$:
$$
\Phi =(-a,\A_1,\A_2,\A_3), \qquad {\rm{curl}} \A = B, \qquad
-\Big(\frac1c\frac{\p \A}{\p t} + \frac{\p a}{\p q} \Big) = E ,
$$
$$
\hat p_0 = -\frac{i\hbar}c\frac{ \p}{\p t} + \frac ec a(t,q), \quad
\hat p_j= -i\hbar\frac{\p}{\p q_j} - \frac ec\A_j(t,q)\quad (j=1,2,3).
\eqno(1.4\rm b)
$$
All the formulae obtained in the paper depend on the symplectic form (1.1)
only, but not on the choice of potentials, and so, all results are gauge
independent.

The noncommutative product $\sf$ which we construct necessarily
reproduces the commutation
relations (1.3) on coordinate functions:
$$
q^j \sf q^k - q^k \sf q^j = 0, \qquad q^j
\sf p_k - q^j \sf p_k = i\hbar \delta^j_k, \qquad
p_j \sf p_k - p_k \sf p_j = i\hbar F_{kj}(q) .
$$
The operators of left multiplication $q\sf$ and $p\sf$ (as well as
right multiplication) are derived without difficulty from
these relations,
using the noncommutative calculus \cite{20,26};
the result incorporates Valatin's \cite{50} primitive of the
closed 2-form $F$ (see Section~2).
We note that the problem of a finding a gauge invariant
symbol product in a convenient closed form was first addressed
by Stratonovich~\cite{54}.

	In Section 3 we derive general representations for $\sf$ in
two equivalent forms, both valid when $F\not= 0$.   The first is
analogous to the exponential Janus derivative representation
\cite{14, 42, 43}
due to Groenewold. The second is a modification of the $(F=0)$
Berezin's integral for the Weyl product.  In both versions there
appears an additional electromagnetic action (flux) over
triangles in phase space.  Moreover, the Groenewold--like
product formula admits generalization for $N$ multipliers via
the magnetic area of polygon membranes.

{}Formulas for the $\sf$ product can be easily
expanded to obtain formal $\hbar\rightarrow 0$ power series.  Higher
order terms beyond the Poisson bracket contribution are functions of
derivatives of $F_{jk}$. This series coincides structurally with
that recently obtained \cite{39} by M\"uller.

The integral formula for $\sf$
is also related to constructions \cite{11, 40, 13} of the Wigner
function in the
presence of electromagnetic fields. In this way in Section~4 we
show that the
$\sf$ product can be produced by a convolution over $T\R^n$.  This
convolution is generated by a version of the Connes' tangential
groupoid \cite{9, 29, 7} but with an additional rapidly oscillating
factor represented by the electromagnetic flux.  Because of the
rapid oscillations, this type of star-product is outside the
framework of the formal deformation quantization method.

In Section 5 we analyze, following the
general approach of \cite{21}, the structure of the symplectic groupoid
corresponding to $\sf$.  This structure is given on the
secondary cotangent
bundle $T^*(T^*\R^n) = T^*\R^n \oplus \R^{2n}$.  The first cotangent
bundle, $T^*\R^n = \R_q^n\otimes\R_p^n$, is the primary phase space,
over which we construct the product $\sf$.  We show how the symplectic
 groupoid structure on $T^*(T^*\R^n)$ senses the magnetic correction $F$ in
the symplectic form, and how the space $\R^{2n}$, dual to $T^*\R^n$, is
equipped with a pseudogroup structure controlled by the Lorentz
momentum of membranes.

{}From this point of view, we claim that in the
formula for $\sf$ it would be more natural to consider not the
usual geodesic triangle, but the triangle with three additional
``wings" directed vertically (i.e., parallel to the
$p$-direction) in the phase space.
The shape of wings is determined by the symplectic groupoid
structure.

        Then in Section~6
we investigate what happens if the Weyl ordering of
noncommuting coordinates is changed to some other ordering.  The
wide (matrix) family of orderings introduced in \cite{28} we relate
to phases in the exponential representation of the $*$-product.
These phases can be again presented as symplectic areas of
membranes.  The membranes are combinations of the basic triangle
with additional wings that are  now not necessarily vertical.
The shape and direction of the wings exactly control the choice
of the ordering in quantization, and again it is related to the
symplectic groupoid structure over the secondary phase space
$T^*(T^*\R^n)$.

The symplectic area of the wings give three additional
contributions to the phase.  The transform from the original Berezin
phase (the Weyl case) to the new one, generated by the wings,
can be considered as a type of gauge transformation of the
``symplectic potential". On the level of $*$-products this is
the transformation from the distinguished Weyl choice to other
ordering choices. In a sense this is the ``gauge" of
quantization.

The special features of the Weyl quantization which have
made it the preferred choice \cite{5, 31, 42, 36, 43} for physical
applications are: 1) it treats $\hat q$ and $\hat p$
symmetrically; 2) self-adjoint operators have
real symbols; and, 3) the Groenewold--Moyal bracket is an even function of
$\hbar$, in particular its leading semiclassical correction is $O(\hbar^2)$,
not $O(\hbar)$.  {}From the symplectic point of view, the Weyl ordering
seems distinguished since the corresponding membranes are of the
simplest shape (no wings).

The Wick normal and anti-normal orderings
$\stackrel{2}{\hat z^*},\stackrel{1}{\hat z}$ and
$\stackrel{1}{\hat z^*},\stackrel{2}{\hat z}$
(where $z=q+ip$) correspond to pure imaginary wings of
membranes in the product formulas.

        Other convenient orderings ---
the standard $\stackrel{2}{\hat q}, \stackrel{1}{\hat p}$
and anti-standard
$\stackrel{1}{\hat q}, \stackrel{2}{\hat p}$ --- correspond to
the case when the wings
are parallel to the basic triangle and the total membrane
becomes a plane rectangle.  These standard and anti-standard
cases correspond to the push- and pull-groupoid structure on
$T\R^n$ (see  Section~4).
{}From the symplectic point of view these cases are singular because
of the totally caustic character of the graph of symplectic groupoid
multiplication corresponding to these cases.

        In the last sections~7 and~8 we apply these ideas to
the quantum dynamical problem: a charged particle in an
electromagnetic field. The basic results for this system
were found by Dirac, Fock, Peierls, Schwinger. Gauge
invariant versions of the WKB approximation were developed
\cite{41, 8, 16, 46, 37, 38} mostly in
terms of integral kernels (Green functions).
In the context of the present paper we can use the $\sf$ symbol
calculus to obtain a phase space gauge invariant treatment of this problem.

  We first study, in Section~7,  the pure magnetic situation without electric
field.  We represent the gauge invariant version of the quantum evolution
equation over phase space
and show how the Marinov phase and summation rule are
generalized in the presence of the magnetic field. Also we investigate how
the formulae for semiclassical solutions sense the generic wings
of the membranes  (i.e., arbitrary, not Weyl ordering).

At the end of Section~7, using the membrane generalization of the
Groenewold formula, we represent the symbol of the evolution operator
exactly in a continual form. This continual membrane
formula is dual to the Feynman path integral representation.

        Then, in Section~8, we consider a time-dependent
electromagnetic field and
represent the gauge invariant quantum equations over phase space in both
the nonrelativistic and relativistic cases. Here we use dynamical
quantum products which are time-dependent. The evolution of
commutation relations in time is controlled by the electric
field.

We describe the semiclassical solution of the Cauchy problem
using membranes in 7-dimensional contact space
$\R_t\times \R^3_q\times \R^3_p$.
The boundary of these dynamic membranes are given by the
solution of two classical systems:  one for the given particle
and an additional one for a ``virtual" particle of infinite
mass.

\setcounter{equation}{0}

\section{Magnetic product for Weyl ordering}

We begin with the definition of magnetic product on a function
space over $\R^{2n}$.
The logic is the following: we transform relations (1.3) to the
standard Heisenberg relations, apply the standard Weyl operators
of regular representation, and then transform back to the magnetic
variables. As the result, we obtain a formula for magnetic
product in terms of left and right regular representation of the
algebra (1.3).

First we recall the properties of closed forms on $\R^n$.

\begin{lem}
Let $F= \frac 1 2 F_{jk}(q) dq^k \wedge dq^j$ be a closed
$2$-form on $\R^n$.
Consider the vector-valued $1$-form $F dq$ with components
$(F(q)\,dq)_j=F_{jk}(q)\,dq^k$ and define the two point vector
potential
\begin{equation}
\label{eq4}
\dst A(q,q') =\frac1{|q-q'|}\int^q_{q'}|\tilde{q}-q'|
F(\tilde{q}) d\tilde q,
\end{equation}
where the integral is taken along the straight line path
from $q'$ to $q$,
and $|\cdot|$ denotes the Euclidean norm on ${\R}^n$.
Then for an arbitrary fixed $q'\in{\R}^n$ the $1$-form $A(q,q')
dq$ is a primitive of~$F$:
$$
d(A\,dq)=F.
$$
The choice of a primitive {\rm(2.1)} {\rm(}gauge choice{\rm)} is
uniquely characterized by the orthogonality condition
\begin{equation}
\label{eq5}
A(q,q')\cdot (q-q') =0.
\end{equation}
\end{lem}

Note that construction (2.1)
is different from that usually  used in proofs of
the known Poincar\'e lemma in the theory of differential forms.
On the other hand, (2.1) is just a simple particular case of the
solution of Lie system related to a general Poisson bracket; (see
\cite{26} page~81 and references therein).
Formula (2.1) was obtained by Valatin
\cite{50} for electromagnetic tensors. The characteristic condition
(2.2) was stressed by Dirac (see \cite{50} p.~101).

We need several other properties of Valatin's primitive.

\begin{lem}
The following formulas hold{\rm:}
$$
d_{q'}(A(q,q')\,dq)=d_q(A(q',q)\,dq'),
$$
$$
A(q,q')-A(q',q)=\int^q_{q'}F(\tilde q)\,d\tilde q,
$$
$$
A(q,q')+A(q',q)=\frac1{|q-q'|}\int^q_{q_m}
|\tilde q-\tilde q^*|\big(F(\tilde q)-F(\tilde q^*)\big)
\,d\tilde q,
$$
where the integrals are taken along the straight line paths, and
$\tilde q^* = 2q_m - \tilde q$ is the point symmetric to $\tilde q$
with respect to the middle point $q_m=\frac12(q+q')$.
\end{lem}

Let $\triangle$ be the triangle in $\R^n$ with vertices
$q,q',q''$. Consider the integral  (flux) of the form $F$:
\begin{equation}
\label{eq5a}
\F{}_{q''} (q,q') \equiv \int_\triangle F.
\end{equation}
Note that here and everywhere in the sequel the orientation of a
membrane corresponds to the sequence of its vertices (or sides)
read from right to left; so the orientation of $\triangle$
corresponds to the sequence $q\leftarrow q'\leftarrow q''$.

\begin{lem}
The following formulas hold{\rm:}
\begin{equation}
\F{}_{q''} (q,q') =\int _{q'}^q A(\tilde q, q'')d\tilde q,
\end{equation}
$$
\frac {\partial}{\partial q} \F{} _{q''} (q,q') =A(q,q'')-A(q,q'),
$$
where the integral in {\rm(2.4)} is taken along the straight line path.
\end{lem}

Now let us fix the second argument of $A(q,q')$ at some point,
say, $q'=0$ and introduce the operators
$$
\hat p'=\hat p+A(\hat q,0).
$$
Since $\hat q$, $\hat p$ satisfy relations (1.3), the new set of
operators $\hat q$, $\hat p'$
satisfy the standard Heisenberg commutation relations
\begin{equation}
[\hat q^j, \hat p'_k] =i\hbar \delta_k^j,
\quad [\hat q^j, \hat q^k] = [\hat p'_j, \hat p'_k] =0.
\end{equation}

Any Weyl-symmetrized function of operators  $\hat q,\hat p$
can be transformed to a function of operators  $\hat q,\hat p'$
by the formula
\begin{equation}
f(\hat q,\hat p   )=f \bigg(\frac {\stackrel {3} {\hat q} +
\stackrel {1} {\hat q} } 2 ,\,
\stackrel {2} {\hat p'} -
{\tilde A} (\stackrel {3} {\hat q},\stackrel {1} {\hat q}) \bigg),
\end{equation}
where
$$
{\tilde A}(q,q') \equiv \int _0^1 A(q\mu+q'(1-\mu), 0)d\mu.
$$
We have used here formulas of noncommutative analysis \cite{20} and
\cite{26} pp.~277--295; the superscripts on top of operators
denote the order of application.

Throughout the paper we will not give a characterization of the
spaces the symbols must belong to in order that (2.6) and the subsequent
product formulas are well defined.  This is a separate technical (and
often not simple) question which has been extensively investigated in the
pseudodifferential operator literature \cite{10,18,28,26}.
The reader can
consider all formulas as formally algebraic or, depending on the formula,
assume an appropriate simple symbol class such as polynomials, smooth
rapidly decreasing functions, etc.

{}For the Heisenberg algebra (2.5) the operators of left and
right regular representation are well known.
Namely, consider arbitrary Weyl-symmetrized
function $g'$ with operator arguments $\hat q,\hat p'$:
\begin{equation}
\hat g' \equiv g'(\hat q,\hat p') =
g' \bigg(\frac {\stackrel {3} {\hat q} +
\stackrel {1} {\hat q} }2 ,\, \stackrel {2} {\hat p'} \bigg).
\end{equation}
Then  the following left and right multiplication formulas hold
\cite{10,26,28,2}:
\begin{equation}
\begin{array}{ll}
\hat q \hat g' = \widehat{L_q'g'}, \quad & L_q' =q+\frac 1 2 {i\hbar}
 \partial_{ p'};\\[2ex]
\hat p' \hat g' = \widehat
{L_p'g'}, \quad & L_p' =p'-\frac 1 2 {i\hbar}
 {\partial_ q};\\[2ex]
\hat g' \hat q = \widehat{R_q'g'}, \quad & R_q' =q-\frac 1 2 {i\hbar}
\partial_{ p'};\\[2ex]
\hat g' \hat p' = \widehat{R_p'g'}, \quad & R_p' =p'+\frac 1 2 {i\hbar}
 {\partial_ q}.
\end{array}
\end{equation}
Here we denote $\pa_q\equiv \pa/\pa q$ and
$\pa_{p'}=\pa/\pa p'$.
Operators $L',R'$ satisfying multiplication formulas (2.8) are
called the left and right regular representation of the given
algebra, in our case, the algebra (2.5).

On the right-hand side of the formula (2.6)
operators  $\stackrel {1} {\hat q}$ and $\stackrel {3} {\hat q}$
in arguments of $\tilde A$ can be considered as multiplication
by $\hat q$ from the left and from the right,
and so they can be replaced by $L_q'$ and $R_q'$ acting on
arguments of $f$. Thus we obtain from (2.6)
$$
f(\hat q,\hat p   )=f' \bigg(\frac {\stackrel {3}
{\hat q} + \stackrel {1} {\hat q}}2,\, \stackrel {2} {\hat p'} \bigg)
=f'(\hat q,\hat p'),
$$
where
$$
f'(q,p') =f(q,p'-{\tilde
A}(\overleftarrow {L_q'},\overleftarrow {R_q'}))=
\exp \{ -\tilde A (L_q',R_q') \partial _{p'}\}f(q,p'),
$$
and the left arrows mean that  operators act on arguments
standing to their left.
After substitution of  explicit formulas for $L_q'$,
$R_q'$ from (2.8) we conclude
$$
f'(q,p')  = \exp \{ -{\tilde A}
(q+{\textstyle\frac12}{i\hbar} {\partial_{ p'}},
q-{\textstyle\frac12}{i\hbar} \partial_{ p'})
\partial_{ p'} \}f(q,p').
$$

Note that by Lemma~2
$$
{\tilde A} (q+ u/ 2, q- u/ 2 ) u = \int _{q-u/2}^{q+u/2}
A(\tilde{q},0) d\tilde{q} =\F{}_0 (q+ u/ 2,q-u /2).
$$

So we obtain the transformation formula

\begin{prop} 
Any Weyl-symmetrized function $f$ in operators $\hat q,\hat p$,
satisfying commutation relations {\rm (1.3)} , can be transformed to
the Weyl-symmetrized function $f'$    in operators $\hat q,\hat p'$
satisfying Heisenberg relations {\rm(2.5)}. This transform is given
by formula
\begin{equation}
f'=U_F f, \quad U_F = \exp
\Big\{\frac i \hbar \F{}_0 (q+{\textstyle\frac12} {i\hbar} \partial_p,
q-{\textstyle\frac12} {i\hbar}\partial_p)\Big\},
\end{equation}
where the  $\F{}_0$ was defined in {\rm(2.3)}.
\end{prop}

Using this transform one easily obtains all objects which are
needed for the algebra (1.3). For instance, the operators of
left and right regular representation for (1.3) are the
following
$$
\begin{array}{ll}
L_q = U_F^{-1} \cdot L_q'\cdot U_F,\quad
&R_q = U_F^{-1} \cdot R_q'\cdot U_F,\\[2ex]
L_p = U_F^{-1} \cdot L_p'\cdot U_F-A(L_q,0),\quad
&R_p = U_F^{-1}\cdot  R_p'\cdot U_F-A(R_q,0).
\end{array}
$$

Applying explicit formulas (2.8) for $L',R'$, (2.9) for
$U_F$ and using Lemma~3 we get
\begin{equation}
\begin{array}{ll}
L_q =q +\frac12{i\hbar} \partial_p,\quad
& R_q =q -\frac12{i\hbar} \partial_p,\\[2ex]
L_p =p -\frac12{i\hbar} \partial_q-A(L_q,R_q),\quad
& R_p =p +\frac12{i\hbar} \partial_q-A(R_q,L_q).
\end{array}
\end{equation}

\begin{lem}
Operators $L_q$, $L_p$ satisfy commutation
relations {\rm(1.3)}, and operators $R_q$, $R_p$ satisfy the
conjugate relations {\rm(}with opposite signs{\rm)}.
Operators $L$ commute with $R$.
\end{lem}

If we know the regular representation
of commutation relations, we know the product of symbols
(see \cite{26} Appendix 2).

\begin{theor}
The product of Weyl-symmetrized functions in
operators satisfying relations {\rm(1.3)} is given by
\begin{equation}
f(\hat q,\hat p) \cdot g(\hat q,\hat p)= (f\sf g)(\hat q,\hat p),
\end{equation}
where $\sf$ is the associative product of functions over $\R^{2n}$
defined by
$$
f\sf g = f(L_q,L_p) g =g(R_q,R_p)f.
$$
\end{theor}

We call $\sf$ the {\it magnetic product } of Weyl type.
In particular,  we have
$L_q=q\sf$, $L_p=p\sf$, and $R_q=\sf q$, $R_p= \sf p$.
These multiplication operators satisfy relations (1.3)
and their conjugate companions;
left and right multiplications commute with each other.

\setcounter{equation}{0}

\section{Exponential formula for magnetic product}

The next useful stage in evaluation of the $\sf$-product
is to bring it into an exponential form. First,
we apply Proposition 1 and the usual Groenewold formula \cite{14,10}
known for Weyl product of symbols over the Heisenberg algebra,
namely,
\begin{equation}
f'\star g'
= f' \exp \Big\{- \frac{i\hbar}2 \ola{\partial} J^{-1}\ora{\partial}\Big\}g'
=f'(q+ \ora {u}/2,p)\,\, g'(q- \ola{u}/2, p),
\end{equation}
where
$$
\ola{u} = i\hbar \ola{\partial_p},\quad \ora{u} = i\hbar \ora{\partial_p},
\quad \partial=(\partial_ q, \partial_ p).
$$

{}From (2.9) we calculate
$$
\begin{array}{ll}
 U_F f\star U_F g &\dst = f\exp \left\{
\frac i\hbar \F{}_0(q+ (\ola u +\ora u)/2, q+ (\ora u -\ola
u)/2 )\right.\\ [2ex]
&\dst \left.+ \frac i\hbar \F{}_0(q+ (\ora u -\ola u)/2, q-
(\ola u +\ora u)/2) -\frac {i\hbar}2 \ola\partial J^{-1}
\ora\partial\right\} g.
\end{array}
$$

Applying the inverse transformation $U_F^{-1}$ we obtain
$$
\begin{array}{ll}
U_F^{-1} (U_F f \star U_F g) = &\dst f \exp\left\{
-\frac i\hbar \F{}_0(q+ (\ola u +\ora u)/2, q- (\ola u +\ora
u)/2)\right.\\ [2ex]
&\dst\left.\right. +\frac i\hbar \F{}_0(q+ (\ola u +\ora u)/2,
q+ (\ora u -\ola u)/2)  \\[2ex]
 &\dst\left.+ \frac i\hbar
\F{}_0(q+ (\ora u -\ola u)/2, q- (\ola u +\ora u)/2) -\frac
{i\hbar}2 \ola\partial J^{-1}\ora\partial\right\} g.
\end{array}
$$

\bigskip\noindent
{\small $\lefteqn{\mbox{\unitlength=1pt
\begin{picture}(108.000 ,108.000 )
\put(20,0){Figure 1.}
\put( 24.000 , 27.000 ){$q$}
\put( 70.000 , 95.000 ){$0$}
\put( 61.000 , 49.000 ){$u_1$}
\put( 73.000 , 27.000 ){$u_2$}
\end{picture}}}
{\mbox{\includegraphics{mypic1.ps}}}$ }
\bigskip

The sum of three fluxes which appeared at last exponent can be simplified
if we look at the geometrical picture.
Indeed, the three fluxes represent integrals of the form $F$ over
three triangles with common vertex $0$. They are sides of the tetrahedron.
By the Stokes theorem these three fluxes together are equal to the flux
over the bottom triangle; see Fig.~1, in which the vectors $u_1$ and $u_2$
represent the operators $\ora{u}$ and $\ola{u}$, respectively.

The relation
$$
f\sf g = U_F^{-1} (U_F f\star U_F g)
$$
and the representation of its right hand side which we have
derived above generate the following statement.

\begin{prop}
The magnetic product $\sf$, corresponding to commutation relations
{\rm(1.3)}, can be calculated by the formula
\begin{equation}
(f\sf g)(q,p) = f(q,p) \exp \left\{ \frac i \hbar
\phi (q, i\hbar \ola
{\partial_ p}, i\hbar \ora{\partial_ p})
+\frac{i\hbar}2 \left(
\ola{\partial_ q} \ora{\partial_ p}-
\ola{\partial_ p} \ora{\partial_ q}\right)
\right\} g(q,p
).
\end{equation}
Here
\begin{equation}
\phi(q,u_2, u_1) =\int _{\triangle_q( u_2, u_1)}  F,
\end{equation}
and $\triangle_q( u_2, u_1)$ is a membrane in $\R^n$ whose boundary is
the triangle constructed by the middle point $q$ of one side and by the two
other sides $u_2,u_1$. The magnetic flux {\rm(3.3)}
can also be represented in the form
$$
\phi=\int^1_0d\mu\int^\mu_0d\nu\,\,
u_2F\big(q+(\mu-{\textstyle\frac12})u_1
+(\nu-{\textstyle\frac12})u_2\big)u_1.
$$
\end{prop}

Note the differential operators in the exponent (3.2) act only on the
arguments of target functions $f$ and $g$, but not on the argument $q$ in
the flux $\phi$.

Also note that operators $i\hbar \partial_p\,$, which we
substitute into the flux $\phi$ in the exponent (3.2),
are of order $O(\hbar)$ over the
space of non-oscillating functions as $\hbar\to 0$. So the flux
$\phi$ actually is of
order $O(\hbar^2)$, and the right hand side of (3.2) can be easily
expanded as a power series in $\hbar$. Thus for non-oscillating (as
$\hbar\to 0$) functions $f,g$
\begin{multline}
f\sf g\simeq\sum^\infty_{|\gamma|,|\varepsilon|=0}
\frac{(-1)^{|\varepsilon|}}{\gamma!\varepsilon!}
\Big(-\frac{i\hbar}{2}\Big)^{|\gamma|+|\varepsilon|}\,\\
\times
\partial^\varepsilon_q\partial^\gamma_p f
\exp\bigg\{
-\frac{i\hbar}{2}\sum^\infty_{|\alpha|,|\beta|=0}
\frac{c_{|\alpha|,|\beta|}}{\alpha!\beta!}
\Big(-\frac{i\hbar}{2}\Big)^{|\alpha|+|\beta|}\,
 (\ola{\partial_ p})^\alpha
\langle \ola{\partial_ p}, \partial^{\alpha+\beta} F(q)\ora{\partial_ p}
\rangle (\ora{\partial_ p})^\beta \bigg\}
\partial^\gamma_q\partial^\varepsilon_p g,\nonumber
\end{multline}
where
$$
c_{s,m}=\frac{e[m+1]}{(s+1)(m+1)}
-\frac{e[s+m]}{(s+1)(s+m+2)}\,,
$$
and $e[l]=0$ if $l$ is even, and $e[l]=1$ if $l$ is odd. Here
$\langle {\bf\cdot},\bf{\cdot} \rangle$ is the Euclidean scalar
product in $\R^n$ and the greek letters are multi-indices.

        Of course, the first two terms of this
expansion are $$ f\sf g =fg -\frac {i\hbar}2 \{ f,g\}_F +O(\hbar^2), $$ where
$\{,\}_F$ is the Poisson bracket on $\R^{2n}$ corresponding to the symplectic
form $\omega_F$, i.e.,
\begin{equation} 
\{f,g\}_F = \partial_p f\partial_q g- \partial_q f\partial_p g
+\langle\partial_p f, F(q)\partial_p g \rangle.
\end{equation}
In the case where $F$ is restricted to be the pure magnetic form
(1.2a), then the $\hbar$
series above is equivalent to the one derived in \cite{39}.

        The formula (3.2) is still not presented in a completely symplectic
manner.  We have there the flux $\phi$, which is the integral of the form
$F$ over the triangle in $q$-space. The other part of the exponent in
(3.2) can also be related to the area of a triangle, but in the
$(q,p)$-space; see Fig.~2

\bigskip\noindent
{\small \input pic2.tex }
\bigskip

This triangle $\Sigma_x( V_2, V_1) $ is constructed by the middle
point $x$ of one side, and by the two opposite sides
$ V_2$, $ V_1$. Its
projection onto $\R^n_q$ coincides with the triangle
$\triangle_q(u_2, u_1)$.

\begin{theor}
The magnetic product can be represented in the form
\begin{equation}
(f\sf g)(x) =
\exp \left\{ \frac i \hbar \int_{\Sigma_x(\widehat V_2,\widehat V_1)}
\omega_F\right\}f(x_2) g(x_1)\Big|_{x_1=x_2=x},
 \end{equation}
where $\omega_F$ is the magnetic symplectic form {\rm(1.1)},
$\widehat V=(i\hbar\p_p,i\hbar\p_q)$,
and  $x=(q,p)\in\R^{2n}$.
\end{theor}

In (3.5) one has to evaluate first the symplectic
$\omega_F$-area of the membrane $\Sigma_x(V_2,V_1)$,
and then substitute the operators $\widehat{V}_1, \widehat{V}_2$
for the vectors $V_1,V_2$;
the operator $\widehat{V}_1$ is applied to the argument~$x_1$ and
$\widehat{V}_2$ to the argument~$x_2$.

Actually, formula (3.5) may be immediately generalized to the case
of several multipliers. Consider a membrane in $\R^{2n}$ whose
boundary is formed by $N$ vectors $V_1,V_2,\dots,V_N$  as sides of a
polygon, and by additional point $x$ as a middle point of the $(N+1)$th
side. Denote this membrane by $\Sigma_x(V_N,\dots, V_1)$; see Fig.~3.

\bigskip\noindent
{\small $\lefteqn{\mbox{\unitlength=1pt
\begin{picture}( 96.000 , 96.000 )
\put(20,0){Figure 3.}
\put( 31.000 , 26.000 ){$x$}
\put( 73.000 , 14.000 ){$V_1$}
\put( 85.000 , 55.000 ){$V_2$}
\put( 22.000 , 78.000 ){$V_N$}
\end{picture}}}
{\mbox{\includegraphics{mypic4.ps}}}$ }
\bigskip

\begin{cor}
The following formula holds{\rm:}
\begin{equation}
(f_N\sf \dots \sf f_1)(x) =\exp\bigg\{\frac i \hbar \int_
{\Sigma_x(\widehat V_N,\dots, \widehat V_1)} \omega_F \bigg\}
f_N(x_N)\cdots f_1(x_1)\Big|_{x_1=\cdots=x_N=x}.
\end{equation}
\end{cor}

An immediate consequence of (3.5) is an integral
formula for $f\sf g$. Indeed, in (3.5) we have a
pseudodifferential operator acting on $f$ and $g$. By the usual formulas
via Fourier and inverse Fourier transform we easily calculate this action
and obtain an integral representation for $f\sf g$.
The triangle $\Sigma_x( V_2,  V_1)$ in this new formula will be
described not by the sides $ V_2, V_1$,
but by the middle points $x_2,x_1$ of
those sides. Such a triangle in $\R^{2n}$, constructed by three
middle points $x,x_2,x_1$, we denote by $\Sigma (x,x_2,x_1)$.

\begin{prop}
The magnetic product of the Weyl type is
given by the integral formula
\begin{equation}
(f\sf g)(x) =\frac 1{(\pi\hbar)^{2n}} \int_{\R^{2n}} \int_{\R^{2n}}\
\exp \bigg\{ \frac i \hbar \int_{
\Sigma(x,x_2,x_1)}\omega_F\bigg\} f(x_2)g(x_1)\,dx_2\,dx_1\,.
\end{equation}
\end{prop}

Actually, the
magnetic part $F$ of the form $\omega_F$ is integrated in
(3.5) and (3.7) over the projection of $\Sigma (x,x_2,x_1)$ onto
$q$-space.

In the case $F=0$, (3.7) becomes the Berezin formula for the
Weyl product, and (3.6) becomes equivalent to the $N$-factor
Weyl products obtained in \cite{42, 43}.

\setcounter{equation}{0}

\section{Tangential groupoid and magnetic cocycle}

{}Formula (3.7) for magnetic noncommutative product over
$\R^{2n}=T^*\R^n$ can also be derived  from the
Connes type tangential groupoid structure on $T\R^n$ \cite{9,29,7},
equipped with an additional magnetic phase factor (cocycle).

The space $T\R^n$ consists of pairs $(q,u)$, where $q\in \R^n$ and $u\in
T_q\R^n$. Vector $u$ is interpreted as a ``replacement"
at point $q$. The first and
simplest groupoid structure on $T\R^n$  is given by formula
$(q_2,u_2)\circ (q_1,u_1)= (q_1,u_2+u_1)$  iff $ q_1=q_2$.
This groupoid is commutative; it is called the {\it Galileo groupoid}.
This structure is shown in Fig.~4.
The set of units of this groupoid consists of all pairs $(q,0)$;
so, the set of units is the configuration space $\R^n_q$
considered as the zero-section in $T\R^n$.

\bigskip\noindent
{\small $\lefteqn{\mbox{\unitlength=1pt
\begin{picture}( 98.000 , 98.000 )
\put(20,0){Figure 4.}
\put( 19.000 , 34.000 ){$u_2$}\put( 51.000 , 85.000 ){$u_2+u_1$}\put( 63.500 , 41.000 ){$u_1$}\put( 53.000 , 15.000 ){$q_2=q_1=q$}\end{picture}}}{\mbox{\includegraphics{mypic6.ps}}}$ }
\bigskip

Other geometric combinations of the vectors $u_1, u_2$
and the points $q_1, q_2$ may be assembled to give
noncommutative groupoids. Three simple possibilities are shown
in Fig.~5.

\bigskip\noindent
{\small \input pic5.tex }
\bigskip

\noindent
The multiplication rule, for instance, in the {\it pull-groupoid}
is
$(q_2,u_2)\circ (q_1,u_1)= (q_1,u_2+u_1)$
iff $q_1+u_1=q_2$.
{}For the  {\it saddle-groupoid} one has
$$
(q_2,u_2)\circ (q_1,u_1)= (q,u_2+u_1) \quad {\rm iff}\quad
q_1+u_1/2=q_2-u_2/2.
$$
In this case, $q$ is the middle point of the third side of the
triangle:
$q=q_1+u_2/2=q_2-u_1/2$.

All these groupoids are specific cases of a general
$\tau$-{\it groupoid}, where $0\leq \tau\leq1$.
Pull-, push-, and saddle-cases correspond to $\tau=1$, $\tau=0$,
and $\tau=\frac12$, respectively.
In Fig.~6 one can see the generic case corresponding to some
$\tau$ \  $(\frac12<\tau<1)$; in this case
$q_2=q_1+\tau u_1+(1-\tau)u_2$ and $q=q_1+(1-\tau)u_2$.
The set of units for all of these groupoids is
$\R^n_q\subset T\R^n$.

\bigskip\noindent
{\small $\lefteqn{\mbox{\unitlength=1pt
\begin{picture}(160.000 ,104.000 )
\put(20,0){Figure 6.}
\put(52,22){$q_1$}
\put(71,45){$u_1$}
\put(70,70){$q_2$}
\put(51,85){$u_2$}
\put(30,32){$q$}
\end{picture}}}
{\mbox{\includegraphics{mypic7.ps}}}$ }
\bigskip

On any (measurable) groupoid there is a convolution of distributions
\cite{9, 44}
\begin{equation}
(\phi_2\ostar \phi_1)(a) =
\int_{a=b\circ c} \phi_2(b)\phi_1(c) \,d\mu_b(c),
\end{equation}
where $d\mu_b$ is the Haar measure on fibres of the left
groupoid mapping $c\to c\circ c^{-1}$.
 For example, in the case of saddle-groupoid we have
\begin{equation}
(\phi_2\ostar \phi_1)(q,u)= \int_{\R^n}
\phi_2 (q+u_1/2,u-u_1)\phi_1(q-(u-u_1)/2,u_1)du_1.
\end{equation}

Each convolution over $T\R^n$ generates a noncommutative product over
$T^*\R^n$ just by Fourier transform between $u$ and $p$ coordinates:
\begin{equation}
f^{\sim} (q,u) \equiv
\frac 1 {(2\pi\hbar)^n} \int _{\R^n}e^{iup/\hbar} f(q,p)\,dp=
\frac 1 {(2\pi)^n} \int _{\R^n}e^{iuk} f(q,\hbar k)\,dk.
\end{equation}

{}For example, if one takes the saddle-groupoid convolution
$\ostar$ (4.2) then the corresponding product
over $T^*\R^n$ is the usual Groenewold-Moyal product $\star$
(3.1), i.e.,
\begin{equation}
(f\star g)^{\sim}= f^{\sim}  \ostar g^{\sim}.
\end{equation}

\begin{rem}
\rm
In this approach the quantum convolution $\ostar$ is not an
$\hbar$-deformation of a commutative one. The deformation
parameter $\hbar$
appears only in the Fourier transform (4.3);
this is what  makes it possible to
include into the quantization scheme functions over $T^*\R^n$
that oscillate rapidly as $\hbar \to 0$.
For non-oscillating functions $f(q,p)$ the
transform (4.3) admits the classical limit as $\hbar \to0$;
here one obtains
the correspondence $f(q,p)\to f(q,0)\delta(u)$, $\hbar\to 0$ between the
classical commutative algebra of functions over the phase space $T^*\R^n$
and the commutative subalgebra of distributions over $T\R^n$ of the type
$\varphi(q)\delta(u)$ concentrated at the zero ``replacement"
$u=0$.  Any replacement by a vector $u_0$, represented over $T\R^n$ as the
distribution $\delta(u-u_0)$, after the transform inverse to (4.3)
generates the phase space observable $f(q,p)=\exp\{-iu_0 p/\hbar\}.$
The corresponding operator $\hat f$  (if the magnetic tensor $F=0$)
is just the Heisenberg translation operator
$\hat f \psi (q) = \psi (q-u_0)$.
This is the simplest example of rapidly oscillating functions $f$ which
one has to include into algebra of observables over $T^*\R^n$,
but which lie outside the formal deformation quantization
approach; see also \cite{29, 45}.
\end{rem}

Now let us return to the relation (4.4) between star-product and
convolution.
The question arises: how to get the magnetic product $\sf$
(3.7) via the groupoid structure on $T\R^n$?
Let us slightly generalize (4.1) by introducing an exponential phase factor:
\begin{equation}
(\phi_2\ostar \phi_1)(a) = \int_{a=b\circ c}
e^{i\Phi(b,c)} \phi_2(b)\phi_1(c) \,d\mu_b(c).
\end{equation}
Here $\Phi$ is a groupoid cocycle, i.e.,
$$
\Phi(d,b\circ c)-\Phi(d\circ b,c)+\Phi(b,c)-\Phi(d,b)=0,\quad
\Phi(b,c)=-\Phi(c^{-1}, b^{-1})
$$
for any $b,c,d\in T\R^n$.
The cocycle $\Phi$ is called the coboundary iff
$$
\Phi(b,c) =\psi(b)+\psi(c)-\psi(b\circ c), \quad
\psi(b^{-1})=-\psi(b).
$$
If $\Phi$ in (4.5) is a coboundary, then convolutions (4.5) and
(4.1) are actually equivalent.

In the case of saddle-groupoid we can take the following
``magnetic'' cocycle (coboundary):
\begin{equation}
\Phi=\frac 1 \hbar \int_{\triangle(q,q_2,q_1)} F,
\end{equation}
where $F$ is the Faraday 2-form over $\R^n$, and
$\triangle(q,q_2,q_1)$ is the triangle with middle points $q,q_2,q_1$.
In this case the coboundary function $\psi$ at the point $b=(q,u)$
is given by the integral $\psi(b)=\frac1\hbar\int \widetilde{A}\,d\tilde q$
along the chord $[q-\frac12u, q+\frac12u]$.
Note that $\psi$ controls the additional phase factor in the gauge
invariant version of the Wigner function~\cite{40, 13}.

\begin{prop} 
Let $\sff$ be the saddle-groupoid convolution, equipped with the
magnetic cocycle {\rm(4.6)}.
Then the Fourier transform {\rm (4.3)} relates this convolution
to the magnetic product $\sf$ via
$$
(f\sf g)^{\sim} = f^{\sim} \sff g^{\sim}.
$$
\end{prop}

Note that other types of groupoids (like pull-, push- ) will also
generate some magnetic products over $\R^{2n}=T^*\R^n$. In
Section 6 we will consider a
variety of such products. On the other hand, we will show that not all of
products over $T^*\R^n$ are generated from the convolution over $T\R^n$ in
this fashion.

In Section 5 we analyze the magnetic product (3.7) from the
view point of symplectic groupoid structure of the secondary phase space
(on the secondary cotangent) $T^*(T^*\R^n)$. Regarding this we mention the
following.

\begin{prop}
Each groupoid structure on $T\R^n$ whose
set of units is $\R^n_q$, uniquely determines a symplectic
groupoid structure on $T^*(T^*\R^n)$. Here $T^*\R^n $ is equipped with
the symplectic form $\omega_0 =dp\wedge dq$. If, in addition, on $T\R^n$ a
cocycle of type {\rm(4.6)} is given, then on $T^*(T^*\R^n)$ we have
a symplectic groupoid structure corresponding to the magnetic form
$\omega_F$.
\end{prop}

\setcounter{equation}{0}

\section{Symplectic groupoid and membranes with wings}

{}Formula (3.7) represents the exact magnetic product. On the other hand,
it is known \cite{26}, in a
very general context, how to construct quantum
products of functions over an arbitrary Poisson manifold ${\cal N}$ in the
semiclassical approximation, to all orders in $\hbar\to 0$. This
approximate product takes the form
\begin{equation}
(f*g) (x) \simeq \int\int K_\hbar (x,x_2,x_1)f(x_2)g(x_1) \,dx_1\,dx_2.
\end{equation}
Here $x,x_2,x_1 \in {\cal N}$, and $K_\hbar$ is a ``wave
function" corresponding to some Lagrangian submanifold
$\Lambda_*$ in the ``phase space" $\E \times \E\times \E$, where
$\E$ is the symplectic
groupoid over ${\cal N}$. In our case ${\cal N} = \R^{2n}
=T^*\R^n$ with Poisson
bracket (3.4). The Lagrangian submanifold $\Lambda_*$ is the graph
of groupoid multiplication in $\E$. If this graph is one-to-one
projected onto the ``configuration" space ${\cal N}\times{\cal
N}\times{\cal N}$ along the polarization then the function
$K_\hbar$ is just a WKB function
\begin{equation}
K_\hbar = \exp\{ iS/\hbar\}\varphi +O(\hbar)
\end{equation}
whose phase $S$ is the Poincar\'e-Cartan action on $\Lambda_*$
and $\varphi$ is the solution of the corresponding transport
equation.

In specific cases, for instance, in our case ${\cal N}=\R^{2n}$
with bracket (3.4), the asymptotic formula (5.1) becomes exact
and the remainder $O(\hbar)$ in (5.2) is absent. Indeed, let us
compare (5.1), (5.2) with formula
(3.7) for the magnetic product $\sf$. First, one can describe the
symplectic groupoid $\E$ in our specific case.
Let us set $\E =T^*\R^{2n} =\R_x^{2n} \oplus\R^{2n}_y$,  and denote by
$(x,y)$  points in $\E$. The space $\R_x^{2n}$ is imbedded into $\E$ as a
zero section $\{y=0\}$. The symplectic form on $\E$ is $dy\wedge dx$.  The
operators $L,R$ (2.10) of left and right regular representation of algebra
(1.3) generate two mappings
\begin{equation} 
l:\E \to \R^{2n}, \quad r
:\E \to\R^{2n}.
\end{equation}
Here $l=(l_q,l_p)$ and $r=(r_q,r_p)$ are just
symbols of $L=(L_q,L_p)$ and $R=(R_q,R_p)$, i.e.,
$$
L=l(x, -i\hbar\partial_x), \quad R=r(x, -i\hbar \partial_x).
$$
In our case formula (2.10) reads
\begin{equation}
\begin{array}{ll}
l_q(x,y) =x_q - y_p/2, \quad & l_p(x,y) =x_p+ y_q/2-A(l_q,r_q),\\[2ex]
r_q(x,y) =x_q + y_p/2, \quad & r_p(x,y) =x_p-y_q/2-A(r_q,l_q).
\end{array}
\end{equation}

Mappings (5.3) are Poisson and anti-Poisson, i.e., $l$ preserves
brackets, $r$ changes the sign of brackets (recall that on $\R^{2n}$ we
have the bracket (3.4), and the bracket on $\E$ corresponds to the
symplectic form $dy\wedge dx$).

The groupoid structure on $\E$ is defined as follows: points $m_2,m_1 \in
\E $ are called {\it multiplicable} iff $r(m_2)=l(m_1)$; the product
$m=m_2\circ m_1$, by definition, is a point in $\E$ such that
$l(m) =l(m_2)$, $r(m)=r(m_1)$.
The subspace $\R^{2n}_x\subset{\cal E}$ is the set of units of
this groupoid, and mappings (5.3) are left and right reduction
mappings:
$$
l(m)=m\circ m^{-1},\qquad r(m)=m^{-1}\circ m.
$$

The graph $\Lambda_* \subset
\E\times\E\times\E$ of this groupoid multiplication consists of all
multiplicable points and their products
\begin{equation}
\Lambda_* =\{(m,m_2,m_1) | m=m_2\circ m_1\}.
\end{equation}
If on $\E\times \E\times \E$ we introduce the symplectic form
$dy\wedge dx-dy_2\wedge dx_2-dy_1\wedge dx_1$, then the
submanifold $\Lambda_*$ is Lagrangian (see details in \cite{21, 51}).

In the case $F=0$ (i.e., A=0) formulas (5.4) are interpreted as
``middle point of chord" relations:  $x=\frac12(l+r)$, $y=JV$, where
$V=l-r\in
T\R^{2n}$.
So, the groupoid structure on $\R^{4n}=T^*\R^{2n}$ is
given by the triangle rule (see Fig.~7):
\begin{equation}
\begin{array}{l}
m=m_2\circ m_1, \quad m=(x,y), \quad m_1=(x_1,y_1),\quad m_2=(x_2,y_2);
\\[2ex]
y=y_1+y_2,\qquad
x=x_1+{\textstyle\frac12}J^{-1}y_2
 =x_2-{\textstyle\frac12}J^{-1}y_1.
\end{array}
\end{equation}

\bigskip\noindent
{\small \input pic7.tex }
\bigskip

{}For arbitrary given triple of points $x, x_2,x_1\in \R^{2n}$ we uniquely
construct the triangle for which these points are the middle points of
its sides, and so reconstruct elements $m,m_2,m_1\in \E$ such that
$m=m_2\circ m_1$. This means that the graph $\Lambda_*$ is one-to-one
projected onto $\R^{2n}\times \R^{2n}\times \R^{2n}$
along the ``vertical" $y$-polarization.
Hence, the kernel $K_\hbar$ has the WKB form (5.2),
and its phase is
\begin{equation}
S(x,x_2,x_1)=\int_{(0,0,0)}^{(x,x_2,x_1)} (ydx-y_2dx_2-y_1dx_1).
\end{equation}
Here
$y,y_2,y_1$ are determined via $x,x_2,x_1$ following the triangle
multiplication rule; the initial point $(0,0,0)$ corresponds to triple of
elements $m_0=m_0\circ m_0$, where $m_0=(0,0) \in \E$.

\bigskip\noindent
{\small $\lefteqn{\mbox{\unitlength=1pt
\begin{picture}(170.000 ,120.000 )
\put(20,0){Figure 8.}
\put(14,20){$0$}
\put(150,28){$x$}
\put(133,46){$x_2$}
\put(160,68){$x_1$}
\end{picture}}}
{\mbox{\includegraphics{mypic9.ps}}}$ }
\bigskip

The integral (5.7) is taken over an arbitrary path on
$\Lambda_*$
connecting the triple $(0,0,0)$ (i.e., the degenerate triangle)
with the
triple $(x,x_2,x_1)$ (i.e. the given triangle). This path is actually a
family of triangles in $\R^{2n}$; see Fig.~8. The phase
(5.7) in the case $F=0$ is just equal to the area of the final
triangle
$$
S(x,x_2,x_1)= \int _{\triangle(x,x_2,x_1)} \omega_0.
$$

Now let us see what happens in the general magnetic case $F\ne 0$.
{}From (5.4) it follows that the groupoid structure now is given by
formulas
\begin{equation}
x={\textstyle\frac12} ({l+r}) +V_F,\quad y=JV+Y_F,\quad \mbox{where}\quad
V=l-r.
\end{equation}
Here we have introduced notations
\begin{equation}
\begin{array}{l}
V_F= ( 0; A^s)\in \R^n_q\oplus \R^n_p, \quad
Y_F= ( A^a;0)\in \R^n_{y_q}\oplus \R^n_{y_p},\\[2ex]
A^
s=\frac 1 2 (A(l_q,r_q)+A(r_q,l_q)),\quad
A^a= A(l_q,r_q)-A(r_q,l_q).
\end{array}
\end{equation}

Comparing with the multiplication rule (5.6) in the case $F=0$
we see that points $x,x_2,x_1$ are no longer the middle points
of sides of the  basic triangle and do not even belong to those
sides. They are shifted in the $p$-direction by the vectors $V_F$.

Another difference from the case $F=0$ is that the usual
multiplication rule
$y=y_1+y_2$ (see (5.6)) fails to hold for inhomogeneous magnetic
case. New ``magnetic" rule is
\begin{equation}
y=y_1+y_2+( \int_{\triangle} \nabla F ;\,\, 0),
\end{equation}
where the vector-valued closed $2$-form $\nabla F$  is defined by
$\nabla F = \frac12 \nabla F_{jk} dq^k \wedge dq^j$, and
$\triangle=\triangle (q,q_2,q_1)$ is the triangle in $q$-space
with middle points
$(q,q_2,q_1)$, which are projections of $x,x_2,x_1\in \R^{2n}$
onto $\R^n_q$. Note that
$$
-\int_\triangle \nabla F = \oint_{\partial \triangle} F\,dq \,\,\sim\,\,
\mbox{\rm Lorentz momentum}.
$$
The vector 2-form $\nabla F$ is a measure of {\it magnetic
inhomogeneity}. So,
this form controls the modification of the usual ($F=0$)
multiplication rule.
The space $\R^{2n}_y$ (dual to $\R^{2n}_x$) now is not
even a group, it is a pseudogroup over the Poisson manifold
$\R^{2n}_x$ (see details in general case
in \cite{26}, the word ``pseudo''- reflects the fact that the product
(5.10) in the $y$-space depends on $x$-coordinates as additional
parameters, and the condition of ``associativity'' of the
product (5.10) senses this dependence).
The non-trivial part of the pseudogroup structure (5.10) is
determined by the  momentum of the membrane $\triangle$
in the magnetic field.

{}Formula (5.7) in the case $F\ne 0$ still
represents the phase in the product
(3.7) if we put there the corrected values $y=JV+Y_F$. But now the
triangle no longer represents the groupoid multiplication rule, and
from this point of view, it seems unnatural to keep this triangle in
the formula for the phase.

\bigskip\noindent
{\small \input pic9.tex }
\bigskip

Actually, we have another configuration (see Fig.~9) related to
the triple $(x,x_2,x_1)$. This  configuration consists of triangle
$\Sigma(\tilde x,\tilde x_2,\tilde x_1)$, where $\tilde x =x -V_F$, and
also of three
additional triangles directed ``vertically'' (parallel to $p$)
with ``top'' vertices $x,x_2,x_1$. We call them {\it magnetic wings}.

A magnetic wing is characterized by a sequence of its
vertices $[l,x,r]$ related to each other by (5.8) (see Fig.~10, left picture).
The configuration of membrane with wings, as we see, exactly
corresponds to the new groupoid multiplication rule for $F\ne 0$.

\bigskip\noindent
{\small \input pic10.tex }
\bigskip

\begin{lem}
$$
\int_{\Sigma(\tilde x, \tilde x_2,\tilde x_1)} \omega_F=
\int_{\Sigma( x,  x_2, x_1)} \omega_F,\quad
\int_{{\rm vertical\,\, wing}} \omega_F=0.
$$
\end{lem}

Both statements of the Lemma follow from the orthogonality condition
(2.2).

Denote by $\Sigma_F (x,x_2,x_1)$ the triangle
$\Sigma(\tilde x, \tilde x_2,\tilde x_1)$ together with three
magnetic wings described above. The boundary
of this figure consists of six straight line segments. The figure itself
looks like inflected hexagon. More generally, $\Sigma _F$ could be
any membrane in the phase space $\R^{2n}$ with that six segment boundary.
In view of Lemma~5 we have
$$
\int_{\Sigma( x,  x_2, x_1)} \omega_F=\int_{\Sigma_F( x,  x_2,
x_1)} \omega_F.
$$

\begin{theor}
The Weyl type magnetic product over $T^*\R^n$ is given by
\begin{equation}
(f\sf g)(x) =\frac 1{(\pi\hbar)^{2n}} \int \int
\exp \bigg\{ \frac i \hbar \int_{
\Sigma_F(x,x_2,x_1)}\omega_F\bigg\} f(x_2)g(x_1)\,dx_2\,dx_1,
\end{equation}
where $\Sigma_F$ is the wing membrane corresponding to the
magnetic groupoid structure {\rm(5.8)} on $T^*(T^*\R^n)$.
\end{theor}

\setcounter{equation}{0}

\section{Ordering in quantization}

Now we demonstrate that membrane wings introduced in previous section are
actually very natural objects in the quantization framework. We show how
different configurations
of wings relate to different choices in the
ordering problem. In particular, we'll see why the Weyl ordering choice
(Weyl symmetrization) looks like an optimal choice.

Let us take a constant real $2n\times 2n$ matrix $M$ (actually, a linear
operator in the space tangent to $T^*\R^n$) which obeys the
condition
\begin{equation}
M^T J+JM=0,
\end{equation}
where $M^T$ is the transposed matrix.
One can represent $M$ by its $n\times n$ blocks as follows
$$
M=\left( \begin{array}{ll} N&K\\S&-N^T\end{array} \right).
$$

Now let us take an arbitrary function $f=f(q,p)$
(say, a polynomial) and determine the following general ordering
of operators $\hat q, \hat p'$ (generators of the Heisenberg
algebra):
\begin{equation}
\hat f^M\equiv f\Big({\textstyle\frac12} N(
\stackrel 4{\hat q}-\stackrel 2{\hat q})+
{\textstyle\frac12} ({\stackrel 2{\hat q}+\stackrel 4{\hat q}})
-{\textstyle\frac12} K
(\stackrel 5{\hat p'}-\stackrel 1{\hat p'}),\,\,
\stackrel 3{\hat p'}
-{\textstyle\frac12} S
(\stackrel 4{\hat q}-\stackrel 2{\hat q})
-{\textstyle\frac12} N^T(\stackrel 5{\hat p'}-\stackrel 1{\hat p'})
\Big).
\end{equation}

We would like to obtain the product formula
\begin{equation}
\hat f^M \cdot \hat g^M =\hat k^M, \quad
k=f\stackrel {M} {*}g.
\end{equation}

To derive the product $\stackrel {M} {*}$ we first note that
the $M$-ordering (6.2) is related to the Weyl ordering via
\begin{equation}
\hat f^M=(U^Mf)(\hat q,\hat p'),
\end{equation}
where
$$
U^M =\exp \bigg\{ \frac i \hbar S^M (\widehat V )\bigg\},\qquad
\widehat V=(i\hbar\p_p,i\hbar\p_q),
$$
and the function $S^M$ is defined by the matrix $M$ as follows
$$
S^M (V) =\frac 1 2 \langle JMV, V\rangle
=\int _{\triangle^M (V)} \omega_0.
$$
Here $\triangle^M(V)$ is the triangle, called the $M$-{\it wing},
generated by vector $V$ and by vector $MV$ applied at the middle
point of $V$ (see Fig.~10).

Using $U^M$ we can calculate the product (6.3)
by the  formula
$$
f\stackrel {M} {*} g \equiv U^{-M} (U^M f\star U^M g),
$$
where $\star $ is the
Weyl product (without magnetic correction,
at first).  Looking at exponential representation of the Weyl
product, one concludes that
$$
f\stackrel {M} {*} g  = \exp \Big\{\frac i \hbar \Phi^M (\widehat V_2,
\widehat V_1)\Big\} f(x_2)g(x_1)\Big|_{x_1=x_2=x},
$$
where $\widehat V =(i\hbar\p_p, i\hbar\p_q)$.
In this formula we use the notation
$$
\Phi^M (V_2,V_1) = \int _{\Sigma (V_2,V_1)} \omega_0
+S^M(V_1)+S^M(V_2)-S^M(V_1+V_2),
$$
where
$\Sigma (V_2,V_1)$ is the triangle generated by vectors $V_1$, $V_2$.

The phase function $\Phi ^M$
can be written as the symplectic area of the {\it wing membrane}
$\Sigma_x^M (V_2,V_1)$ generated by four triangles:
$$
\Sigma_x^M (V_2,V_1) =\triangle (V_2, V_1)
\bigcup\triangle^M (V_1) \bigcup
\triangle^M (V_2) \bigcup \triangle^M (V_1+V_2).
$$
Its boundary consists of six line segments.
For reasons of uniformity in notation
the point~$x$ is shown;
actually, the $\omega_0$-area  of $\Sigma_x^M (V_2,V_1)$
is independent of~$x$.

The picture for $\Sigma^M_x$ is the same as in Fig.~9, but wings now are
$M$-wings as in Fig.~10 (right picture).

\begin{theor}
For an arbitrary $2n\times 2n$ matrix $M$,
satisfying {\rm(6.1)}, there is a
star-product over $\R^{2n}$ given by the wing membranes{\rm:}
\begin{equation}
(f\stackrel {M} {*} g)(x)  = \exp \Big\{\frac i \hbar
\int _{\Sigma_x^M(\widehat V_2,\widehat V_1)} \omega_0\Big\} f(x_2) g(x_1)
\Big|_{x_1=x_2=x}.
\end{equation}
This product corresponds to the $M$-ordering rule {\rm(6.2)} of
non-commutative operators $\hat q,\hat p'$ {\rm(}generators of
the Heisenberg algebra{\rm)}, so that formula {\rm(6.3)} holds.
\end{theor}

The family of orderings (6.2) was introduced and studied in detail
in \cite{28}, where the following pseudodifferential formulas for
$M$-product were obtained:
$$
(f\stackrel{M}\star g)(x)
=f\Big(\stackrel{2}x-i\hbar({\textstyle\frac12}-M)J^{-1}
\stackrel{1}{\pa_x}\Big)g(x)
=g\Big(\stackrel{2}x+i\hbar({\textstyle\frac12}+M)J^{-1}
\stackrel{1}{\pa_x}\Big)f(x).
$$

In the particular case
\begin{equation}
M=({\textstyle\frac12}-\tau)
\left( \begin{array}{ll} I&0\\0&-I\end{array} \right)
\end{equation}
the ordering (6.2) simplifies to
\begin{equation}
\hat f^\tau =f\Big(\tau \stackrel 1{\hat q} +(1-\tau) \stackrel 3{\hat q},\,\,
\stackrel 2{\hat p'}\Big).
\end{equation}
The family of $\tau$-wings corresponding to this family of
orderings is represented in Fig.~11.
The whole membrane $\Sigma^M_x$ for the specific case
$\tau=0$ is shown in Fig.~12.

\bigskip\noindent
{\small \input pic11.tex \hbox to 3cm{} \input pic12.tex }
\bigskip


The cases $\tau =0$ and $\tau=1$ are called  the standard and
anti-standard ordering choices (they correspond to push- and
pull- groupoid
structures on
$T\R^n$, see Section 4), and the case $\tau =\frac 1 2 $ is the Weyl
ordering (corresponding to the saddle-groupoid structure on $T\R^n$). In
the later case $M=0$ and wings in membranes $\Sigma^M$ are absent.

Matrix $M$ is assumed to be real, but if we formally take
$M=iJ/2$ (i.e., $N=0$, $S=-i/2$, $K=i/2$),
then
the transform $U^M=\exp\{-\hbar(\pa^2_q+\pa^2_p)/4\}$ relates
Weyl ordering to the Wick normal ordering choice.
In this case wings are pure imaginary (see Fig.~13), and the
membrane $\Sigma^M$ coincides with hexagon membrane introduced
in~\cite{22, 23}.

\bigskip\noindent
{\small $\lefteqn{\mbox{\unitlength=1pt\begin{picture}(110.000 ,110.000 )\put(20,0){Figure 13.}\put( 12.000 , 80.000 ){$l$}\put( 80.000 , 12.000 ){$r$}\put( 84.000 , 55.000 ){$\partial/\partial z^*$}\put( 45.000 , 88.000 ){$\partial/\partial z$}\end{picture}}}{\mbox{\includegraphics{mypic16.ps}}}$ }
\bigskip

\begin{rem}
\rm In the general case, block $N$  of the matrix $M$ controls the
position of the point $q\in \R^n$ with respect to
the middle point of the vector $u$ in the
$T\R^n$-groupoid interpretation.
The block $S$ controls an  additional $T\R^n$-groupoid
cocycle   (in $u$-coordinates) which appears in the convolution
formula (see Section~4).
In contrast to that,
the block $K$ in the matrix $M$ generates the transformation of
distributions over $T\R^n_q$ of the following type:
$\exp \{ -i\hbar  K\p_q\cdot \p_q/2 \}$.
This is not a point transformation. Thus the
product $\stackrel {M}{*}$ corresponding to matrix $M$ with $K\ne 0$
can
not be obtained from the groupoid convolution over $T\R^n$
by construction of Section~4. In particular, the Wick product is
of such type.
\end{rem}

Now let us consider the symplectic groupoid structure on
the secondary phase space $T^*(T^* \R^n)$
corresponding to the $\am$ product (6.3). In the same way as in
Section 2 we calculate for the Heisenberg algebra (2.5) operators of
left and right representations corresponding to the ordering choice
(6.2). We know the transformation from (6.2) to the Weyl ordering;
it is given by operator (6.4).
Thus the left operators $L_q^M=q\am$, $L^M_{p'}=p'\am$ and right
operators $R_q^M=\am q$, $\R^M_{p'}=\am p'$ are given by
\begin{equation}
L^M= U^{-M} \cdot L'\cdot U^M, \quad
R^M= U^{-M} \cdot R' \cdot U^M,
\end{equation}
where $L',R'$ are determined in (2.8).
We represent $L^M,R^M$ via their symbols $l,r$:
$$
L^M =l (x,-i\hbar \partial_x),\quad
R^M =r (x,-i\hbar \partial_x),
$$
and easily calculate $l,r$ by (6.8) and by the definition of $U^M$.
The result is
$$
l(x,y) =x+({\textstyle\frac12} -M)J^{-1} y, \quad
r(x,y) =x-({\textstyle\frac12} +M)J^{-1} y.
$$
The inverse mapping $(l,r)\to(x,y)$ is given
by
\begin{equation}
x={\textstyle\frac12}({l+r}) +MV, \quad V \equiv l-r =J^{-1} y.
\end{equation}
For example, for the ordering cases (6.7), formulas (6.9) are
represented in Fig.~11.

The corresponding symplectic groupoid structure on $T^*(T^*\R^n)
=\R_x^{2n} \oplus \R_y^{2n}$ is given by the rule represented
in Fig.~14.
We observe that this structure is exactly given by the membrane with
wings $\Sigma^M$ which we described at the beginning of this section.
What is new now is that we have identified
the positions of points $x,x_2,x_1$
exactly as vertices of wings of the membrane $\Sigma^M$.

\bigskip\noindent
{\small \input pic14.tex }
\bigskip

\begin{cor}
Let $\det(\frac12-M)\ne0$, then the integral version of the
product formula {\rm(6.5)} reads
\begin{equation}
(f\am g )(x) =\frac 1 {(2\pi\hbar)^{2n}|\det(\frac12-M)|}
\int\int
\exp \Big\{ \frac i \hbar \int _{\Sigma ^M(x,x_2,x_1)} \omega_0\Big\}
f(x_2) g(x_1) dx_2 dx_1,
\end{equation}
where $\Sigma ^M(x,x_2,x_1)$ is the membrane with three $M$-wings
having vertices $x,x_2,x_1$.
\end{cor}

Note that for the $\tau$-ordering case (6.6) the denominator in
formula
(6.10) $\det(\frac12-M)=\tau^n(1-\tau)^n$ is zero if $\tau=0$ or
$\tau=1$.
We see that the cases $\tau=0$ (standard ordering) and
$\tau=1$ (anti-standard ordering) are special.
In these cases the graph $\Lambda_*$ of the groupoid
multiplication (see Section~5) is not one-to-one projected onto
the ``configuration space'' $\R^{2n}_x\times\R^{2n}_{x_2}\times
\R^{2n}_{x_1}$, i.e., the graph is totally caustic and the
integral kernel $K_\hbar(x,x_2,x_1)$ (5.1) is not of WKB-type
(5.2).  This is one of essential diffferences between
representations (6.5) and (6.10) for the $*$-product.

\begin{cor}
For several multipliers there is   the following product formula
\begin{equation}
(f_N\am \cdots\am f_1 )(x) =
\exp \Big\{ \frac i \hbar \int _{\Sigma_x
^M(\widehat V_N,\dots,\widehat V_1)} \omega_0\Big\} f_N(x_N)\dots f_1(x_1)
\Big|_{x_1=\cdots =x_N =x}\,\,,
\end{equation}
where $\Sigma_x^M(V_N,\dots,V_1)$ is the membrane with
$N+1$ wings.
\end{cor}

These formulas are naturally generalized for products of differently
quantized multipliers, i.e., for the case when each multiplier has its own
ordering choice:
\begin{equation}
\begin{array}{ll}
\hat f_N^{M_N}\cdot \dots \hat f_1^{M_1}=\hat k^{M_{N+1}},\\[2ex]
 k(x) =\dst
\exp \Big\{ \frac i \hbar \int _{\Sigma_x^{M_{N+1},\dots,M_1}
(\widehat V_N,\dots,\widehat V_1)} \omega_0\Big\} f_N(x_N)\dots f_1(x_1) \Big|_{x_1
=\cdots =x_N =x}.
\end{array}
\end{equation}
Here the membrane $\Sigma_x^{M_{N+1},\dots,M_1}$ is constructed by $N+1$
wings each of different configuration determined by different matrices
$M_{N+1},\dots,M_1$.

\bigskip
In conclusion of this section we consider the magnetic case $F\ne 0$. For
simplicity, we concentrate on the $\tau$-ordering choice
\begin{equation}
\hat f^\tau =f(\tau \stackrel 1{\hat q} +(1-\tau) \stackrel 3{\hat q},\,\,
\stackrel 2{\hat p}),
\end{equation}
where $\hat q$, $\hat p$ satisfy the magnetic commutation relations
(1.3).

We first transform the magnetic $\tau$-ordering to the magnetic Weyl
ordering:
\begin{equation}
\hat f^\tau = (U^\tau f) (\hat q, \hat p), \quad
U^\tau =\exp \{({\textstyle\frac12}-\tau) \hat u \partial_q\},
\end{equation}
where $\hat u =i\hbar \partial_p$. Then for product of two
$\tau$-ordered  observables we have from (2.11):
$$
\hat f^\tau \cdot \hat g^\tau =
(U^\tau f) (\hat q, \hat p)\cdot
(U^\tau g) (\hat q, \hat p) =
(U^\tau f  \sf U^\tau g) (\hat q, \hat p),
$$
where  $\sf$ is the magnetic product (3.2), (3.5),
corresponding to the Weyl ordering. Transforming back  the Weyl symbol
to $\tau$-symbol by the transform $(U^\tau)^{-1}$, we obtain the product
formula
$$
\hat f^\tau \cdot \hat g^\tau = \hat k^\tau,
$$
where
$$
k\equiv f\amf g =(U^\tau)^{-1} (U^\tau f \sf U^\tau g).
$$
By formula (3.2) we calculate the new $\tau$-magnetic product as
follows
\begin{eqnarray}
&&(f \amf g) (q,p) = \\[2ex]
&&\qquad\textstyle\exp \{ (\tau -\frac 1 2 )\hat u \partial_q \}
\Big(\exp \{ \frac i\hbar \phi(q,\hat u_2,\hat u_1)+
\frac 1 2 (\hat u_1 \partial_{q_2} -\hat u_2 \partial _{q_1})\}
\nonumber\\[2ex]
&&\qquad\textstyle\times\exp \{(\frac 1 2-\tau )
(\hat u_1 \partial_{q_1} +\hat u_2 \partial _{q_2})\}
f(q_2,p_2)g(q_1,p_1)\Big)\Big|_{q_1=q_2=q,\,p_1=p_2=p}
\nonumber\\[2ex]
&&\qquad\textstyle=\exp \Big\{\frac i\hbar \phi\Big(q+(\tau-\frac 1 2)
(\hat u_1 +\hat u_2),
\hat u_2, \hat u_1\Big) +\frac 1 2 (\hat u_1\partial _{q_2} -\hat u_2\partial
_{q_1})\nonumber\\[2ex]
&&\qquad\textstyle+(\tau -\frac 1 2)
\Big((\hat u_1+\hat u_2)(\partial_{q_1} +\partial_{q_2}) -
\hat u_1 \partial_{q_1} -\hat u_2 \partial_{q_2}\Big) \Big\}
f(q_2,p_2)g(q_1,p_1)\Big|_{q_1=q_2=q,\,p_1=p_2=p}.\nonumber
\end{eqnarray}

Note that the flux $\phi(q,u_2,u_1)$ was defined by (3.3)
via the triangle $\triangle_q(u_2,u_1)$ in $\R^n$ with the
middle point $q$  of one of sides. Now we see that the position
of the middle point $\tilde q$ differs from $q$ by an additional
vector $(\tau -1/2)(u_1+u_2)$.
So,
$$
q=\tilde q +({\textstyle\frac 1 2} -\tau) (u_1+u_2) =\tilde q +(MV)_q,
$$
where $M$ is matrix (6.6) corresponding to $\tau$-ordering, and
$V=V_1+V_2$ is a vector in $\R^{2n}$ whose $q$-component is $u=u_1+u_2$.

The total phase in (6.14) is equal to the magnetic area of the
membrane in $\R^{2n}$ constructed by the triangle with sides
$V_1,V_2,V_1+V_2$ and by three additional  wings generated by
matrix $M$ of special type (6.6).
The configuration of the wings in this special case
is shown in Fig.~11.

 The position of the vertex of the wing is $x^\tau =\tilde x +MV$,
where $\tilde x =(l+r)/2$. But actually this position should be changed
to match the symplectic groupoid  multiplication rule.
To find this rule we have to calculate the operators of left and
right regular
representation in our $\tau$-case. In view of (6.14) these
operators are given by formulas
$$
L^\tau =(U^\tau)^{-1} \cdot L\cdot U^\tau, \quad
R^\tau =(U^\tau)^{-1} \cdot R\cdot U^\tau,
$$
where $L,R$ are operators (2.10), corresponding to the magnetic Weyl
ordering. {}From (6.14) and (2.10) we obtain
$$
\begin{array}{rlrl}
L_q^\tau &=q+\tau i\hbar \partial_p, &
R_q^\tau &=q-(1-\tau) i\hbar \partial_p,\\
L_p^\tau &=p-(1-\tau) i\hbar \partial_q -A(L_q^\tau, R_q^\tau), &
R_p^\tau &=p+\tau i\hbar \partial_q -A(R_q^\tau, L_q^\tau).
\end{array}
$$

So, if we represent these operators by symbols
$L^\tau = l(x, -i\hbar \partial_x)$,
$R^\tau = r(x, -i\hbar \partial_x)$, then two groupoid  mappings
appear
$$
l:\, T^*\R^{2n}\to \R^{2n},\qquad
r:\, T^*\R^{2n}\to \R^{2n},
$$
where
\begin{equation}
\begin{array}{ll}
l(x,y) =x+(\frac 1 2  -M)J^{-1} y -\big(0;A(l_q, r_q)\big),\\
r(x,y) =x-(\frac 1 2  +M)J^{-1} y -\big(0;A(r_q, l_q)\big).
\end{array}
\end{equation}
{}From (6.16) we reconstruct $x$ and $y$ via $l,r$
\begin{equation}
x= {\textstyle\frac12} (l+r) +V_F +
MV, \quad y = JV +Y_F,
\end{equation}
where $V=l-r$, and vectors $V_F, Y_F$ are given by the same
formulas as in (5.9) (i.e., the same as in the case $\tau=1/2$
or $M=0$).

Relations (6.17) determine the final configuration of the {\it magnetic
$\tau$-wing}. Position of the vertex  $x$ of this magnetic wing
is shifted by vector $V_F$ with respect to position $x^\tau$.

\begin{lem}
The magnetic area of the magnetic $\tau$-wing with the vertex $x$ is
equal to the magnetic area of the wing with vertex  $x^\tau$.
\end{lem}

In view of this lemma the phase (6.15) can be represented by the
area of a membrane $\Sigma_x^\tau (V_2, V_1)$ constructed by
the triangle of vectors $V_1,V_2,V_1+V_2$,
and by three magnetic $\tau$-wings over each of these vectors.
The point~$x$ is the vertex of the magnetic $\tau$-wing over
$V_1+V_2$.

\begin{theor}
The magnetic star-product corresponding to $\tau$-ordering
{\rm(6.13)} of  noncommutative coordinates $\hat q,\hat p$ is
given by formula
$$
(f\amf g )(x) = \exp\Big\{ \frac i \hbar \int_{\Sigma_x^\tau(\widehat V_2,
\widehat V_1)}\omega_F \Big\} f(x_2) g(x_1) \Big|_{x_1=x_2=x},
$$
where $\Sigma^\tau_x$ is a membrane with magnetic $\tau$-wings.
\end{theor}

The immediate corollaries from this statement are formulas for several
multipliers and also the integral formula for $\amf$ via the magnetic area
of membranes $\Sigma^\tau (x,x_2,x_1)$ supplied with magnetic
$\tau$-wings. These corollaries are formulated by the same way
as (6.10)--(6.12), but with magnetic form $\omega_F$ in the
exponent.

\setcounter{equation}{0}

\section{Dynamics via membrane area}

Quantum dynamics of a charged (spinless) particle in
electromagnetic field can be described

-- in the nonrelativistic case by the Schr\"odinger equation
\begin{equation}
i\hbar\frac{\p\psi}{\p t}=\Big[\frac1{2m}
\Big(-i\hbar\frac{\p}{\p q}-\frac ec\A\Big)^2+ea\Big]\psi,
\end{equation}

-- in the relativistic case by the Klein--Gordon equation
\begin{equation}
\Big(i\hbar\frac{\p}{\p t}-ea\Big)^2\psi
=c^2\Big[\Big(-i\hbar\frac{\p}{\p q}-\frac ec\A\Big)^2
+m^2c^2\Big]\psi,
\end{equation}
where $\A$ and $a$ are magnetic and electric potentials of the
field, and metric is assumed to be Euclidean.

First, we consider the pure magnetic time-independent situation
$a=0$, $\A=\A(q)$. Let us introduce operators
$\hat p=-i\hbar{\p}/{\p q}-\frac ec\A(q)$.
Then the dynamical equations can be reduced to studying the
evolution operator $\exp\{-{it}H(\hat p)/\hbar\}$,
where $H(p)=p^2/2m$ in the nonrelativistic case, and
$H(p)=\pm c\sqrt{p^2+m^2c^2}$ in the relativistic case.

More generally, in a presence of an additional non-Euclidean
metric
(gravitational field) the Hamiltonian will depend on the
$q$-coordinate as well:
$H=H(q,p)$.
Say, $H\simeq g^{jk}(q)p_jp_k/2m$ in the nonrelativistic case.

So, the general problem is to study the operator
\begin{equation}
U^t=\exp\Big\{-\frac{it}{\hbar}\widehat H\Big\},\qquad
\widehat H=H(\hat q,\hat p).
\end{equation}
In particular, we are interested in its asymptotic behavior as
$\hbar\to0$.

Note that the symbol $H$ and the permutation relations between
quantum coordinates $\hat q, \hat p$ (1.3) are independent of
the gauge choice of the magnetic potential $\A$.
Thus the semiclassical approximation for $U^t$ as $\hbar\to0$,
written in terms of the phase space symbol $H(q,p)$ and the
phase space noncommutative structure (1.3),
is automatically gauge invariant.

Let us represent the operator $U^t$ in the Weyl form
$$
U^t=\U^t(\hat q,\hat p),
$$
then for symbol $\U^t$ one obtains the following equations
\begin{equation}
i\hbar\frac{\p \U^t}{\p t}=H\sf\, \U^t,\qquad \U^0=1.
\end{equation}
Using Theorem 1 we transform  (7.4) to a pseudodifferential
form:
\begin{equation}
i\hbar\frac{\p \U^t(x)}{\p t}=\H_\hbar(x,-i\hbar{\p_x})\,\U^t(x),
\qquad \U^0=1.
\end{equation}
Here $\H_\hbar(x,y)$ is the Weyl symbol of the operator
$H(L_q,L_p)$, and $L_q$, $L_p$ are given by (2.10). Obviously,
\begin{equation}
\H_\hbar=\H_0+O(\hbar^2),\qquad
\H_0\equiv H(l),
\end{equation}
where $l$ is the Weyl symbol of $L$, i.e.,
$L=l(x,-i\hbar\p_x)$.
The explicit formulas for $l=(l_q,l_p)$ are found in (5.4).

In view of (7.6) the principal term of the semiclassical
solution of the Cauchy problem (7.5) is determined
by the Hamilton function $\H_0$. For small enough time interval
$t\in[0,T]$ the approximate solution has the simplest WKB-form
\begin{equation}
\U^t(x)=\exp\Big\{\frac i\hbar S(t,x)\Big\}u^t(x)+O(\hbar),
\end{equation}
where the phase $S$ is the solution of the Hamilton--Jacobi
equation
\begin{equation}
\frac{\p S}{\p t}+\H_0\Big(x,\frac{\p S}{\p x}\Big)=0,\qquad
S\Big|_{t=0}=0,
\end{equation}
and the non-oscillatory amplitude $u^t$ is the solution of the
``transport'' equation
\begin{equation}
\frac{\p u^t}{\p t}+\frac{\p}{\p x}
\bigg(\frac{\p\H_0}{\p y}\Big(x,\frac{\p S}{\p x}\Big)u^t\bigg)=0,
\qquad u^0=1.
\end{equation}

In order to solve (7.8), (7.9) one has to consider the Hamiltonian
system
\begin{equation}
\dot X=\frac{\p \H_0}{\p Y},\qquad
\dot Y=-\frac{\p \H_0}{\p X},\qquad
X\Big|_{t=0}=x^0,\qquad
Y\Big|_{t=0}=0,
\end{equation}
and determine
\begin{equation}
S=\int^t_0(Y\dot X-\H_0)\,d\tilde t,\qquad
u^t=\Big(\frac{\D X}{\D x^0}\Big)^{-1/2},
\end{equation}
where $x^0=(q^0,p^0)$ is taken from the equation
$$
X(x^0,t)=x.
$$
The time interval $t\in[0,T]$ over which the Jacobian
$\D X/\D x^0\geq\delta>0$ (for any $x^0\in\R^{2n}$) is
exactly the interval where the solution $\U^t$ can be represented
in the WKB form (7.7).

\begin{lem} 
Trajectories of {\rm(7.10)} are given by formulas
\begin{equation}
\begin{array}{l}
X(x^0,t)=\frac12(\gamma^t(x^0)+x^0)+v^t(x^0),\\
Y(x^0,t)=J(\gamma^t(x^0)-x^0)+y^t(x^0),
\end{array}
\end{equation}
where
$$
v^t=\Big(0;{\textstyle\frac12}
\big(A(\gamma^t_q,q^0)+A(q^0,\gamma^t_q)\big)\Big),\qquad
y^t=\Big(A(\gamma^t_q,q^0)-A(q^0,\gamma^t_q);0\Big),
$$
and $\gamma^t=(\gamma^t_q,\gamma^t_p)$ is the trajectory of the
Hamiltonian system corresponding to the function $H$ and the
magnetic Poisson bracket {
\rm(3.4)}
\begin{equation}
\dot \gamma_q=\frac{\p H}{\p p}(\gamma_q,\gamma_p),
\qquad
\dot \gamma_p=-\frac{\p H}{\p q}(\gamma_q,\gamma_p)
-F(\gamma_q)\frac{\p H}{\p p}(\gamma_q,\gamma_p),
\end{equation}
$$
\gamma\Big|_{t=0}=x^0=(q^0,p^0).
$$
\end{lem}

The proof of this lemma follows from (5.8) and from the fact
that $\H_0=H(l)$, and so components of the mapping
$r:\,\R^{2n}_x\oplus\R^{2n}_y\to \R^{2n}$ (5.4) are integrals of
motion for system (7.10), i.e.,
$r(X,Y)=r(x^0,0)=x^0$ is constant in time.

After substitution of (7.12) into (7.11) one obtains the
following result.

\begin{theor}
Let $\hat q,\hat p$ satisfy commutation relations {\rm(1.3)}
with magnetic tensor $F_{kj}$, and $\widehat H=H(\hat q,\hat p)$.
Then for small enough time~$t$ the
semiclassical approximation for the magnetic Weyl symbol of the
evolution operator
$\exp\{-\frac{it}{\hbar}\widehat H\}=\U^t(\hat q,\hat p)$
is given by the formula
\begin{equation}
\U^t={\cal J}^{-1/2}\exp\Big\{\frac i\hbar\int_{\Sigma}\omega_F
-\frac{it}{\hbar}H\Big\}+O(\hbar).
\end{equation}
Here the membrane $\Sigma=\Sigma^t_x$ {\rm(}see Fig.~{\rm15},
left picture{\rm)} is constructed from the piece of the Hamilton trajectory
{\rm(7.13)}, which connects points $x^0$ and $\gamma^t(x^0)$,
and from the magnetic wing
with vertices $[\gamma^t(x^0),x,x^0]$, where
\begin{equation}
x={\textstyle\frac12}\big(\gamma^t(x^0)+x^0\big)+v^t(x^0).
\end{equation}
Here
$$
v^t(x^0)=\Big(0\,\,;{\textstyle\frac12}\big(A(\gamma^t(x^0)_q,q^0)
+A(q^0,\gamma^t(x^0)_q)\big)\Big),
\qquad x^0\equiv(q^0,p^0),
$$
and $A$ is the Valatin primitive {\rm(2.1)}.
The Jacobian ${\cal J}$ in {\rm(7.14)}  is given by
$$
{\cal J}(x^0,t)=\det[{\textstyle\frac12}(I+d\gamma^t(x^0))+dv^t(x^0)],
\qquad {\cal J}\geq\delta>0
\mbox{ for } t\in[0,T],
$$
the Hamilton function $H$ in {\rm(7.14)} is evaluated on the
trajectory, i.e.,  $H=H(x^0)$, and $x^0$ is assumed to be
expressed in terms of $x,t$ via the equation {\rm(7.15)}.
\end{theor}

\bigskip\noindent
{\small \input pic15.tex }
\bigskip

\begin{rem}
\rm
The group property of the family of symbols $\U^t$ over
$\R^{2n}$ reads
\begin{equation}
\U^{t_2}\,\,\sf\,\,\U^{t_1}=\U^{t_2+t_1}.
\end{equation}
In terms of WKB-phase functions (7.14), the identity (7.16)
requires that
\begin{equation}
\int_{_{\textstyle\Sigma^{t_2}_{x_2}}}\omega_F+
\int_{_{\textstyle\Sigma^{t_1}_{x_1}}}\omega_F+
\int_{_{\textstyle\Sigma_F(x,x_2,x_1)}}\omega_F
=\int_{_{\textstyle\Sigma^{t_2+t_1}_{x}}}\omega_F
\end{equation}
(see Fig.~16), where $\Sigma_F(x,x_2,x_1)$ is the hexagon membrane
with magnetic wings, defined at the end of Section~5.
Note that in the case $F=0$ formula (7.17) coincides with the
phase addition rule obtained by Marinov \cite{32}.
In that particular case the magnetic ``anomaly'' $v^t$ in (7.15)
is absent and the magnetic wings of the membranes disappear.
\end{rem}

\bigskip\noindent
{\small $\lefteqn{\mbox{\unitlength=1pt
\begin{picture}(150.000 ,150.000 )
\put(20,0){Figure 16.}
\put(50,20){$x_1$}
\put(90,20){$x$}
\put(125,20){$x_2$}
\put(30,100){$t_1$}
\put(120,100){$t_2$}
\end{picture}}}
{\mbox{\includegraphics{mypic18.ps}}}$ }
\bigskip

Also note that the Hamilton function $H$ could be
time-dependent. In this case the first membrane phase factor in
formula (7.14) is the same, but the second phase factor becomes
$\exp\{-\frac{i}{\hbar}\int^t_0H(\gamma^{\tilde t}(x^0),\tilde t)\,
d\tilde t\}$;
the trajectory $\gamma^t$ is now the solution of system (7.13)
with time-dependent Hamiltonian~$H$.

\begin{rem}
\rm
One can use not only Weyl but any other ordering choice to
represent the evolution operator as a function in coordinates
$\hat q,\hat p$. Then formula (7.14) still holds with membrane
$\Sigma$ constructed by  wings corresponding to the given
ordering choice (see Section~6); equation (7.15) and the
Jacobian ${\cal J}$ are changed following (6.17).
Moreover, in \cite{28} it was proved that using and combining different
orderings it is possible to avoid the difficulty with time
limitations $t\in[0,T]$ where the WKB-approximation works.

{}For the Wick ordering choice the wings are pure imaginary
(see Fig.~13) and the membrane representation (7.14) coincides
with those obtained in~\cite{24}. In this case the Jacobian ${\cal J}$
is never zero and representation (7.14) is global in~$t$.
\end
{rem}

In conclusion of this section we apply formula (3.6) to derive
the symbol $\U^t$ not asymptotically but in an exact continual
form.
Namely, one can use the approximation
$\U^{t/N}=\exp\big\{-\frac{it}{\hbar N}H\big\}+O(N^{-2})$
and obtain the Trotter type formula
$$
\U^t=\lim_{N\to\infty}\exp\Big\{-\frac{it}{\hbar N}H\Big\}
\sf\dots\sf\exp\Big\{-\frac{it}{\hbar N}H\Big\}
$$
($N$ multipliers). Applying (3.6), one derives
\begin{equation}
\U^t(x)=\lim_{N\to\infty}\exp\Big\{\frac i\hbar
\int_{\Sigma_x(\widehat V_N,\dots,\widehat V_1)}
\omega_F\Big\}
\exp\Big\{-\frac{it}{\hbar N}\sum^N_{j=1}H(x_j)\Big\}
\bigg|_{x_1=\dots=x_N=x}.
\end{equation}
Now the question is how to represent this formula in a
continual form.

Note that each vector field $v$ on $\R^{2n}$ and any point
$x\in\R^{2n}$, $t\in\R$ determines a membrane
$\Sigma^{t}_x(v)\subset \R^{2n}$ whose boundary is constructed
from a piece  $\{\Gamma^{\mu}\mid 0\leq \mu\leq t\}$ of the
trajectory of the field $v$ in $\R^{2n}$
and from the magnetic wing
with vertices $[\Gamma^t,x,\Gamma^0]$
(or, the magnetic $\tau$-wing if one wants to use the general
$\tau$-ordering). The integral over the membrane in (7.18) is an
approximation of the integral over $\Sigma^{t}_{x}(v)$ with a
convenient choice of~$v$.

\begin{theor}
The following continual formula
for the symbol of the evolution operator {\rm(7.3)} holds{\rm:}
\begin{equation}
\U^t(x)=\exp\Big\{\frac{i}{\hbar}
\int_{\Sigma^{t}_{x}(\widehat v)}\omega_F\Big\}
\exp\Big\{-\frac{i}{\hbar }\int^{t}_{0}H\big(x(\mu)\big)\,d\mu\Big\}
\bigg|_{x(\mu)\equiv x}.
\end{equation}
Here $\big\{x(\mu)=\big(q(\mu),p(\mu)\big)\mid 0\leq\mu\leq
t\big\}$ are continuous paths in $\R^{2n}$
and
$\widehat v=\big(i\hbar\delta/\delta p(\mu),\break
i\hbar\delta/\delta q(\mu)\big)$
is the variational derivative operator acting on the path functional.
\end{theor}

{}Formula (7.19) is dual to the Feynman path-integral
formula \cite{47, 30, 33, 19, 43}.
The difference between (7.19) and the path integral is the same
as between (3.5) and (3.7).
The known Wick and Hori formulas \cite{53, 17} for the symbol of the
evolution operator (see also generalizations in \cite{27})
are structurally close to (7.19), but use a  different
first exponential factor.
The membrane exponential factor in (7.19) clearly demonstrates
the influence of the magnetic form $\omega_F$ to the quantum
dynamics.

\setcounter{equation}{1}

\section{Electromagnetic fields and space-time membranes}

Let us now consider general time-dependent case, i.e.,
$\A=\A(t,q)$, $a=a(t,q)$ in (7.1), (7.2). We again study the
Cauchy problem for the Schr\"odinger or Klein--Gordon equations.

As a first step one can remove the electric potential $a$
from equations by introducing the new wave function
$\psi\exp\{\frac {ie}\hbar\int^t_0 a\,dt\}$.
After such a transform the magnetic potential $\A$ is replaced
by $\A+c\int^t_0({\p a}/{\p q})\,dt$,
but the electromagnetic tensor $F_{jk}$ (1.2b) remains unchanged.
So, without loss of generality one can assume that
$$
a\equiv0,\qquad E=-c^{-1}\p\A/\p t,\qquad B=\curl\A.
$$

The quantum dynamical equations have the following form:

-- in the nonrelativistic case
$$
i\hbar\frac{\p\psi}{\p t}=H(\hat q,\hat p(t))\psi,\qquad
\mbox{where}\quad H\simeq\frac1{2m}g^{jk}(q)p_jp_k;
\eqno(8.1\rm a)
$$

-- in the relativistic case
$$
\hbar^2\frac{\p^2\psi}{\p t^2}+H^2(\hat q,\hat p(t))\psi=0,\qquad
\mbox{where}\quad H^2\simeq c^2(g^{jk}(q)p_jp_k+m^2c^2).
\eqno(8.1\rm b)
$$
In the latter case the metric $g^{jk}$ is assumed to be
non-negative definite; the symbols $\simeq$ in (8.1\,a,b)
mean that some terms of order $\hbar$, $\hbar^2$ could be added
to the Hamilton function \cite{8, 12, 37, 42, 13, 55}.

In equations (8.1)  for each fixed time $t$ the operators
$\hat p(t)=-i\hbar\p/\p q-\frac ec\A(t,q)$ and $\hat q=q$
satisfy relations (1.3) with time-dependent tensor
$$
F_{jk}(t,q)=\frac ec \epsilon_{kjl} B^l(t,q),\qquad
q\in\R^3,\quad j,k=1,2,3.
$$

The time derivative of the operators $\hat p(t)$ in (8.1) is the
following:
\begin{equation}
\frac{d}{dt}\hat p_j(t)=e E_j(
t,q),\qquad j=1,2,3.
\end{equation}
So we see that the electric field is responsible for ``dynamical
evolution'' of the quantum magnetic algebra (1.3).

Let us introduce two-point electric potential
\begin{equation}
\beta(t,q,q')\equiv\int^{q'}_{q}E(t,\tilde q)\,d\tilde q
\end{equation}
(the integral is taken along the straight line segment), and also
the two-point magnetic potential
\begin{equation}
\alpha(t,q,q')\equiv\frac1{|q-q'|}\int^{q}_{q'}|\tilde q-q'|
B(t,\tilde q)\,d\tilde q.
\end{equation}
We stress that these potentials are different from those used by
Valatin \cite{50} in the time dependent case, since in our present
definitions there is no integration over the time variable.
Time and space are separated because we study the Cauchy problem
in time.

\begin{lem}
The relation holds{\rm:}
$$
-\frac{\p \beta}{\p q}-\frac1c\frac{\pa\alpha}{\p t}=E(t,q).
$$
\end{lem}

Now from (8.2) and from composition formulas (2.11) we obtain
the following statement.

\begin{prop}
{\rm(i)} The time derivative of any Weyl function in quantum
coordinates $\hat q$, $\hat p(t)$ is given by
$$
-i\hbar \frac{d}{dt}f\big(\hat q,\hat p(t)\big)
=f^e\big(\hat q,\hat p(t),t\big),
$$
where
$$
f^e(q,p,t)=e\beta(t,L_q,R_q)f(q,p),
$$
and $L_q=q+\frac12i\hbar\p_p$, $R_q=q-\frac12i\hbar\p_p$ are
operators of the regular representation {\rm(2.10)}.

{\rm(ii)} The composition of two Weyl functions is given by
$$
[f_2(\hat q,\hat p(t))]\cdot[f_1(\hat q,\hat p(t))]
=k(\hat q,\hat p(t)),\qquad
k=f_2(L_q,L_p(t))f_1
$$
where
$L_p(t)=p-\frac12i\hbar\p_q-\frac ec\alpha(t,L_q,R_q)$.
\end{prop}

The solution of the evolution problem (8.1a) has the general form
\begin{equation}
\psi(t,q)=\U^t(\hat q,\hat p(t))\Big(\psi\Big|_{t=0}\Big).
\end{equation}
In view of Proposition~6 equations for symbol $\U^t$ are the
following
\begin{equation}
\Big[-i\hbar\frac{\p}{\p t}
+e\beta(t,L_q,R_q)
+H(L_q,L_p(t))\Big]\U^t(x)=0, \qquad \U^0=1.
\end{equation}
The operator acting on $\U^t$ can be represented (as in (7.5))
via a symbol $\H_\hbar$
over
$\R_t\times\R^{6}_x\times\R^{6}_y$.
In the same way as in (7.6) we have
\begin{equation}
\H_\hbar=\H_0+O(\hbar^2),
\qquad
\H_0(t,x,y)=e\beta\big(t,l_q(x,y),r_q(x,y)\big)
+H\big(l_q(x,y),l_p(t,x,y)\big),
\end{equation}
where $x=(q,p)$, $y=(y_q,y_p)$, and
$$
l_q=q-{\textstyle\frac12} y_p,\qquad
r_q=q+{\textstyle\frac12} y_p,\qquad
l_p=p+{\textstyle\frac12} y_q-{\textstyle\frac ec}\alpha(t,l_q,r_q).
$$
(Of course, here we just re-state identities (5.4) in a new notation.)
As in Section 7, the WKB-solution of (8.6) has the form (7.7),
(7.11), where $(X,Y)$ is now the trajectory of the Hamiltonian
system
\begin{equation}
\dot X=\frac{\p\H_0}{\p y}(t,X,Y),\qquad
\dot Y=-\frac{\p\H_0}{\p x}(t,X,Y),\qquad
X\Big|_{t=0}=x^0,\qquad Y\Big|_{t=0}=0.
\end{equation}

The solution of problem (8.1b) with additional Cauchy data
$\p\psi/\p t\big|_{t=0}=0$ can also be constructed in the form (8.5),
where $\U^t$ satisfies the equations
$$
\Big[i\hbar\frac{\p}{\p t}-e\beta(t,L_q,R_q)\Big]^2\U^t
=H^2(L_q,L_p(t))\U^t,\qquad
\U^0=1,\quad \frac{\p}{\p t}\U^t\Big|_{t=0}=0.
$$
The WKB-approximation has the form
\begin{equation}
\U^t=\frac12\sum_{\pm}\exp\Big\{\frac i\hbar S_\pm\Big\}
u^t_\pm+O(\hbar),
\end{equation}
where the phases $S_\pm$ and amplitudes $u^t_\pm$
correspond (by formulas (7.11)) to the Hamilton function
$\H_0$ of type (8.7) with $\pm$ signs in the definition of~$H$.
The Hamiltonian system (8.8) again plays the basic role.

The difference in  Hamiltonian system (8.8) from the earlier
(7.10) is that function
$\H_0$ in (8.8) now depends on $l_q,l_p$ and on $r_q$
as well.
So, $r_p$ is not an integral of motion for (8.8).
Thus instead of dynamical system (7.13) we get now two
systems:
one for $\gamma=l(t,X,Y)$ and
another for $\lambda=r(t,X,Y)$.
They are the following:
\begin{align}
&\dot \gamma_q=\frac{\p H}{\p p}(\gamma_q,\gamma_p),\\
&\dot \gamma_p=-\frac{\p H}{\p q}(\gamma_q,\gamma_p)
-\frac ec B(t,\gamma_q)\times\dot \gamma_q+eE(t,\gamma_q),
\nonumber\\
\intertext{and}
&\dot \lambda_q=0,\\
&\dot \lambda_p=eE(t,\lambda_q)\nonumber
\end{align}
with one and the same initial condition
$\gamma\Big|_{t=0}=\lambda\Big|_{t=0}=x^0$.

The function $H(q,p)$ has the following form:
$H(q,p)=g^{jk}(q)p_jp_k/2m$ in the nonrelativistic case
and
$H(q,p)=\pm c\sqrt{g^{jk}(q)p_jp_k+m^2c^2}$
in the relativistic case.

Note that  (8.10) is the standard dynamical
system for charged massive particle in the electromagnetic
field.
The additional system (8.11)  can be
interpreted as the dynamical system for a particle of charge~$e$
and mass $m=\infty$. The appearance of this additional ``virtual
particle''  is due to the presence of the electric field~$E$.

The phase $S$ of the WKB-solution is given by (7.11); hence,
\begin{equation}
S(t,x)=\int^t_0Y\dot X\,d\tilde t
-e\int^t_0\beta(\tilde t,\gamma^{\tilde t}_q,\lambda^{\tilde t}_q)
\,d\tilde t
-\int^t_0H(\gamma^{\tilde t})\,d\tilde t.
\end{equation}

\begin{lem}
The following identity holds{\rm:}
\begin{equation}
\int^t_0Y\dot X\,d\tilde t
=\int_{_{\textstyle\Sigma^t_x}}\omega_F
=\int_{_{\textstyle\Sigma^t_x}}\omega_0
+\frac ec \int_{_{\textstyle\widetilde{\Sigma}^t_q}}
B(t,\tilde q)\,d\tilde q\wedge d\tilde q,
\end{equation}
where the membrane $\Sigma^t_x\subset\R^6=T^*\R^3$ is constructed
from the two trajectories $\gamma=\gamma^t$ {\rm(8.10)}
and $\lambda=\lambda^t$ {\rm(8.11)} and from the magnetic wing with
vertices $[\gamma^t,x,\lambda^t]$, where $x=(q,p)$
{\rm(}see Fig.~{\rm15}, right picture{\rm)}.
The projection of $\Sigma^t_x$ onto $\R^3$ is the membrane
$\widetilde{\Sigma}^t_q$ constructed by a piece of the trajectory
$\{\gamma^{\tilde t}_q\,|\, 0\leq \tilde t\leq t \}$
and by the chord $[q^0,\gamma^t_q]$ with middle point~$q$.
\end{lem}

This is the membrane area interpretation of the first term in
(8.12). The second term in view of definition (8.3) can also
be written as two-dimensional area, but in extended space-time:
\begin{equation}
-e\int^t_0\beta\,d\tilde t=
e\int^t_0\,d\tilde t\int^{\gamma^{\tilde t}_{q}}_{q^0}
E(\tilde t,\tilde q)\,d\tilde q
=e\int_{_{\textstyle\widetilde{\Sigma}_{t,q^0}}}E\,dq\wedge dt.
\end{equation}
Here $\widetilde{\Sigma}_{t,q^0}$ is a membrane in
$\R_t\times\R^{3}_q$ whose boundary consists of the
trajectory (the world line)
$\{(\tilde t,\gamma^{\tilde t}) \,|\, 0\leq \tilde t\leq t\} \; +$
the chord $[\gamma^{t},q^0] \; + $
the straight time-segment
$\{(\tilde t,q^0) \,|\, 0\leq \tilde t\leq t\}$.

\bigskip\noindent
{\small $\lefteqn{\mbox{\unitlength=1pt
\begin{picture}(230.000 ,160.000 )
\put(20,0){Figure 17.}
\put(80,90){$x_0$}
\put(50,130){$t=0$}
\put(175,135){$t$}
\put(200,115){$\gamma^t$}
\put(220,95){$x$}
\put(200,40){$\lambda^t$}
\end{picture}}}
{\mbox{\includegraphics{mypic19.ps}}}$ }
\bigskip

Now one can combine (8.13), (8.14) and apply the Stokes theorem
to transform the integration area to be of the most elegant
geometry. Let us denote by $\Sigma_{t,x}$ the membrane
in $\R_t\times\R^6_x$
whose boundary is constructed by the world line of the given
particle
$\{(\tilde t,\gamma^{\tilde t}) \,|\, 0\leq \tilde t\leq t\}$,
the world line of the virtual infinitely heavy particle
$\{(\tilde t,\lambda^{\tilde t}) \,|\, 0\leq \tilde t\leq t\}$,
and also by the magnetic wing with vertices
$[\lambda^t,x,\gamma^t]$ (see Fig.~17).
We refer to $\Sigma_{t,x}$ as a  {\it dynamical membrane}.

\begin{prop}
The WKB-phase of symbol $\U^t$ in {\rm(8.5)}
can be represented as
\begin{equation}
S(t,x)=\int_{_{\textstyle\Sigma_{t,x}}}(\omega_0+F)
-\int^t_0H(\gamma^{\tilde t})\,d\tilde t,
\end{equation}
where $\omega_0=\frac12J dx\wedge dx$,
the $2$-form $F$ is given by {\rm(1.2b)},
$\Sigma_{t,x}$ is the dynamical membrane in
$\R^7=\R_t\times\R^6_x$,
and $\gamma^t$ is the solution of classical dynamical system
{\rm(8.10)}.
\end{prop}

\begin{rem}
\rm
The closed $2$-form $\tilde \omega_F=\omega_0+F$,
which appeared in (8.15), generates a contact structure
on $\R^7=\R_t\times \R^6_x$ \cite{1, 34}.
The ``virtual'' system (8.11) is the characteristic system for
$\tilde \omega_F$. More precisely, the vector field
on $\R^7$ corresponding to (8.11) is
$$
v_0=\frac{\p}{\p t}+eE(t,q)\frac{\p}{\p p},\qquad x=(q,p).
$$
This is the null-field for $\tilde \omega_F$:
$$
v_0\rfloor \tilde \omega_F=0,
$$
and the flow of $v_0$ preserves $\tilde \omega_F$:
$$
\L_{v_0}\tilde \omega_F=0.
$$
Here we denote by $\L$ the Lie derivative and use the sign
$\rfloor$ for the contraction of a vector field and a form:
$v\rfloor\omega\, (u)\equiv \omega(u,v)$ for all~$u$.
If one denotes by $v_H$ the vector field on $\R^7$ corresponding
to (8.10)
$$
v_H=\frac{\p}{\p t}+\frac{\p H}{\p p}(q,p)\frac{\p}{\p q}
+\Big(eE(t,q)-\frac ecB(t,q)\times\frac{\p H}{\p p}(q,p)
-\frac{\p H}{\p q}(q,p)\Big)\frac{\p}{\p p},
$$
then
$$
v_H\rfloor \tilde \omega_F=dH-v_0(H)\,dt,\qquad
\L_{v_H}\tilde\omega_F=d(v_0(H))\wedge dt.
$$
Here $v_0(H)=eE\p H/\p p$; so we see how the electric field $E$
determines the ``nonconservation'' properties of the charged
particle dynamics in the contact space $\R^7=\R_t\times\R^6_x$.
\end{rem}

Now let us return to the WKB-representation (7.7), (7.11) of
symbol $\U^t$ and calculate the  Jacobian ${\cal J}=\D X/\D x^0$.
The trajectory $X$ of system (8.9) is given now by a
modification of (7.12):
$X(x^0,t)={\textstyle\frac12}
\big(\gamma^t(x^0)+\lambda^t(x^0)\big)+v^t(x^0)$,
where $v^t$ is the same as in (7.12). Since the solution
$\lambda=\lambda^t$ of (8.11) is easily calculated:
$\lambda^t_q=q^0$,
$\lambda^t_p=p^0+e\int^t_0E(\tilde t,q^0)\,d\tilde t$,
we derive
\begin{equation}
{\cal J}=\det\bigg[\frac12(I+d\gamma^t)+
\bigg(\begin{array}{cc} 0&0\\C^t&D^t\end{array}\bigg)\bigg].
\end{equation}
Here
$$
C^t\equiv\frac ec\frac{\p}{\p q^0}
\Big(\alpha^s\big(t,\gamma^t_q(x^0),q^0\big)\Big)
+e\int^t_0\frac{\p E(\tilde t,q^0)}{\p q^0}\,d\tilde t,
\qquad
D^t\equiv \frac ec\frac{\p}{\p p^0}
\Big(\alpha^s\big(t,\gamma^t_q(x^0),q^0\big)\Big).
$$
The function $\alpha^s$ is determined by
$\alpha^s(t,q,q')\equiv{\textstyle\frac12}
\big(\alpha(t,q,q')+\alpha(t,q',q)\big)$,
where the two-point magnetic potential $\alpha$
is given by (8.4).
The point $x^0$ everywhere in these formulas has to be expressed
via $t,x$ by means of the equation
\begin{equation}
x={\textstyle\frac12}
\big(\gamma^t(x^0)+\lambda^t(x^0)\big)+v^t(x^0).
\end{equation}
This equation is uniquely solvable while the Jacobian is positive
\begin{equation}
{\cal J}\geq\delta>0,\qquad t\in[0,T].
\end{equation}
So, we conclude with the following result.

\begin{theor}
The symbol $\,\U^t$ of the evolution operator {\rm(8.5)} solving
the equation of motion {\rm(8.1a)} or {\rm(8.1b)} can be represented
{\rm(}for sufficiently small time {\rm(8.18))} in the WKB-form
{\rm(7.7)} or {\rm(8.9)}
over the contact space $\R^7=\R_t\times \R^6_x$.
The phases $S$ are given by membrane formula {\rm(8.15)}
and amplitudes $u^t={\cal J}^{-1/2}$  by {\rm(8.16)}.
\end{theor}

\begin{rem}
\rm
Of course, the contact space $\R^7$ can be symplectified (see
\cite{1, 34}) up to $\R^8=(\R_{p_0}\oplus\R_t)\times\R^6_x$ with
symplectic form $\omega'_F=dp_0\wedge dt+\tilde \omega_F$.
The dynamical membrane $\Sigma_{t,x}$ in (8.15)
can be blown up to a membrane $\Sigma'_{t,x}$
in such a way that the path $\gamma^t$
is put on the level $p_0=-H$, and the path $\lambda^t$ is put on
the level $p_0=0$;
so, the summand $\int^t_0H\,d\tilde t$ in (8.15) is included
into the membrane area, and altogether one obtains
$$
S(t,x)=\int_{\Sigma'_{t,x}}\omega'_F.
$$
\end{rem}

{\bf Acknowledgments}. We are grateful to S.~A.~Fulling for
informing us of references \cite{54,39,7}
and for a number of important comments,
and to F.~H.~Molzahn for critical remarks.
We are also indebted to
M.~Kondrat'eva and M.~Shishkova for useful discussions and help with
preparation of the paper. The first author is grateful to Russian Basic
Research Foundation for partial support (grant No.~99-01-01047). The research
of T.A.O.  is supported by a grant from Natural Sciences and Engineering Research
Council of Canada. The authors thank the Winnipeg Institute of Theoretical
Physics for its continuing support.

\break

\end{document}